\documentclass[aps,prb,twocolumn,showpacs,superscriptaddress,floatfix]{revtex4-1}
\usepackage{dcolumn}
\usepackage{amsmath}
\usepackage{graphicx,epsfig,psfrag}
\usepackage{amssymb}
\usepackage{natbib}
\usepackage{url}
\usepackage[breaklinks=true]{hyperref}
\hypersetup{
        colorlinks=true,
}
\usepackage{bookmark}
\renewcommand{\vec}[1]{{\bf #1}}

\begin{document}

\title{Conductance scaling in Kondo-correlated quantum dots:\\ 
Role of level asymmetry and charging energy}
 
\author{L. Merker}
\affiliation
{Peter Gr\"{u}nberg Institut and Institute for Advanced Simulation, 
Research Centre J\"ulich, 52425 J\"ulich, Germany}
\author{S. Kirchner}
\affiliation
{Max Planck Institute for the Physics of Complex Systems, 01187 Dresden, Germany}
\affiliation
{Max Planck Institute for Chemical Physics of Solids, 01187 Dresden, Germany}
\author{E. Mu\~noz}
\affiliation
{Facultad de F\'isica, Pontificia Universidad Cat\'olica de Chile, Casilla 306, Santiago 22, Chile}
\author{T. A. Costi}
\affiliation
{Peter Gr\"{u}nberg Institut and Institute for Advanced Simulation, 
Research Centre J\"ulich, 52425 J\"ulich, Germany}

\begin{abstract}
The low temperature electrical conductance through correlated quantum dots provides a sensitive probe of the 
physics (e.g., of Fermi-liquid versus non-Fermi-liquid behavior) of such systems. Here, we investigate 
the role of level asymmetry (gate voltage) and local Coulomb
repulsion (charging energy) on the low-temperature and low-field scaling properties of the linear conductance of a quantum dot 
described by the single-level Anderson impurity model. We use the numerical renormalization group 
to quantify the regime of gate voltages and charging energies
where universal Kondo scaling may be observed and also quantify the deviations from this universal behavior with
increasing gate voltage away from the Kondo regime and with decreasing charging energy. We also compare our results 
with those from a recently developed method for linear and non-linear transport, which is based on renormalized 
perturbation theory using dual fermions, finding excellent agreement at particle-hole symmetry
and for all charging energies and reasonable agreement at small finite level asymmetry. Our results could be
a useful guide for detailed experiments on conductance scaling in semiconductor and molecular quantum dots exhibiting
the Kondo effect.
\end{abstract}

% new: diamagnetism and paramagnetism, 75.20.Hr
% new: electronic conduction in metals and alloys, 72.15.Qm
% keep: 71.27.+a Strongly correlated electron systems; heavy fermions
%  73.21.La	Quantum dots
%  73.63.Kv	Quantum dots - electron. transport in nanoscale materials and structures
%Heat capacity of crystalline solids: 65.40.Ba
% entropy, thermodynamics, 05.70.-a
% removed: 73.20.-r Electron states at surfaces and interfaces
%removed: 71.30.+h 	Metal-insulator transitions and other electronic transitions 

\date{\today}

\maketitle

\section{Introduction} 
\label{sec:introduction}

Artificial nanostructures, such as semiconductor quantum dots,
\cite{Goldhaber1998,Cronenwett1998,Potok2007,Kretinin2011} magnetic atoms adsorbed on surfaces, 
\cite{Madhavan1998,Otte2008,Li1998} and, molecules attached to leads, \cite{Park2002,Yu2004,Roch2009,Parks2010,Scott2013}
provide new realizations of the Kondo effect of a local spin interacting antiferromagnetically with conduction electrons.
In contrast to their bulk counterparts, \cite{Hewson1997}
these systems are also highly tunable, for example, via application of gate voltages to modify the energy levels of the quantum dot or molecule, or to tune the
tunnel couplings between the leads and the dot.
In addition, application of a finite transport voltage allows an experimental investigation of the effects of strong
correlations on non-equilibrium transport through these model nanosystems, thereby motivating also the development 
of new theoretical approaches for non-equilibrium . \cite{Hershfield1993,Hettler1998,Oguri2005,Rosch2003a,HanHeary2007,Mehta2008,Anders2008,Ratiani2009,
Pletyukhov2012,Smirnov2013,Aligia2012,Munoz2013,Kirchner2012}

Motivated by recent experiments on conductance scaling in correlated quantum dots exhibiting the Kondo effect,\cite{Grobis2008,Kretinin2011,Scott2013} 
we present in this paper a detailed study of the low-temperature and low-field scaling properties of the linear conductance of a quantum dot
described by the single-level Anderson impurity model. Scaling in physical properties is a
hallmark of the Kondo effect.\cite{Hewson1997} Thus, a Kondo model description of a quantum dot implies
that the conductance $G(T,B)$ is a universal function of $T/T_{0}$ and $g\mu_{\rm B}B/k_{\rm B}T_{0}$ over all temperatures $T$ and magnetic 
fields $B$, with microscopic parameters (such as the Kondo exchange $J$) only entering through the dynamically generated
low energy scale $T_{0}$ (to be defined explicitly in Sec.~\ref{sec:nrg-cond}), with $g$, $\mu_{\rm B}$, $k_{\rm B}$ denoting the $g$-factor, Bohr magneton
and Boltzmann's constant respectively. In particular for $T\ll T_{0}$ or $g\mu_{\rm B}B\ll k_{\rm B}T_{0}$ the 
functions $G(T,B=0)=G(0,0)(1-{c}_{\rm T}(T/T_{\rm 0})^{2})$ and $G(T=0,B)=G(0,0)(1-{c}_{\rm B}(g\mu_{\rm B}B/k_{\rm B}T_{\rm 0})^{2})$ exhibit Fermi liquid
corrections about the unitary conductance $G(0,0)$ with deviations which
are universal in the sense that the coefficients $c_{T}=\pi^{4}/16$ and $c_{B}=\pi^{2}/16$ are independent of microscopic details.
Actual quantum dot devices, however, have a finite charging energy, and they are more realistically described
by an Anderson model. The finite charging energy, and the ability to change the level energy of the quantum 
dot with a gate voltage, allow for charge fluctuations (even in the ``Kondo regime'' of the quantum dot)
and can give rise to deviations from the expected Kondo scaling. It is therefore of some interest 
to quantify the effect of increasing charge fluctuations on the values of $c_{T}$ and $c_{B}$.
Recently, this issue has also been addressed in Ref.~\onlinecite{Munoz2013} by using 
a renormalized perturbation theory on the Keldysh contour \cite{Hewson1993,Oguri2005} 
formulated using dual fermions. \cite{Rubtsov2008,Hafermann2009,Jung2012}. This approach, denoted henceforth as
superperturbation theory (SPT), yields both the linear and non-linear conductance. In
this paper we shall compare the predictions of this theory for the linear conductance 
with results obtained within the numerical renormalization group (NRG) approach.\cite{Wilson1975,KWW1980a,*KWW1980b,Bulla2008}

The outline of the paper is as follows. Section~\ref{sec:model} describes the quantum dot model. Section~\ref{sec:nrg-cond} 
gives a brief description of
the calculation of the finite temperature linear conductance $G(T,B)$ of the Anderson model within the NRG following the procedure
in Ref.~\onlinecite{Yoshida2009}
. In Sec.~\ref{sec:fermi liquid}, some Fermi liquid results for $c_{T}$ and $c_{B}$ are given, and
in Sec.~\ref{sec:SPT} we outline the SPT calculations of $c_{T}$ and $c_{B}$, with which we shall compare. 
In Sec.~\ref{sec:results} we present results for the dependence of the 
coefficients $c_{\rm T}$and $c_{\rm B}$  on charging energy and gate voltage (local level energy). The latter are compared
with the corresponding results from SPT. We conclude in Sec.~\ref{sec:conclusions} with a  discussion of the relevance 
of our results for experiments on quantum dots. In Appendix~\ref{appendix:campo} we give an alternative derivation of the discretization
scheme of Campo {\it et al.} in Ref.~\onlinecite{Campo2005}, which we have used in the NRG calculations reported in this paper. This derivation
is carried out for an energy dependent hybridization function following the procedure in Refs.~\onlinecite{Bulla1997,Bulla2008}.
In Appendix~\ref{appendix:cb calculation}, we provide details of the SPT calculation of $c_{B}$ in terms of
renormalized parameters. The relation between bare and renormalized parameters, required for comparing SPT results for $c_{B}$ and $c_{T}$ 
with the NRG results, is also described in Appendix~\ref{appendix:cb calculation}.

\section{Model}
\label{sec:model}
\begin{figure}
\includegraphics[width=\linewidth,clip]{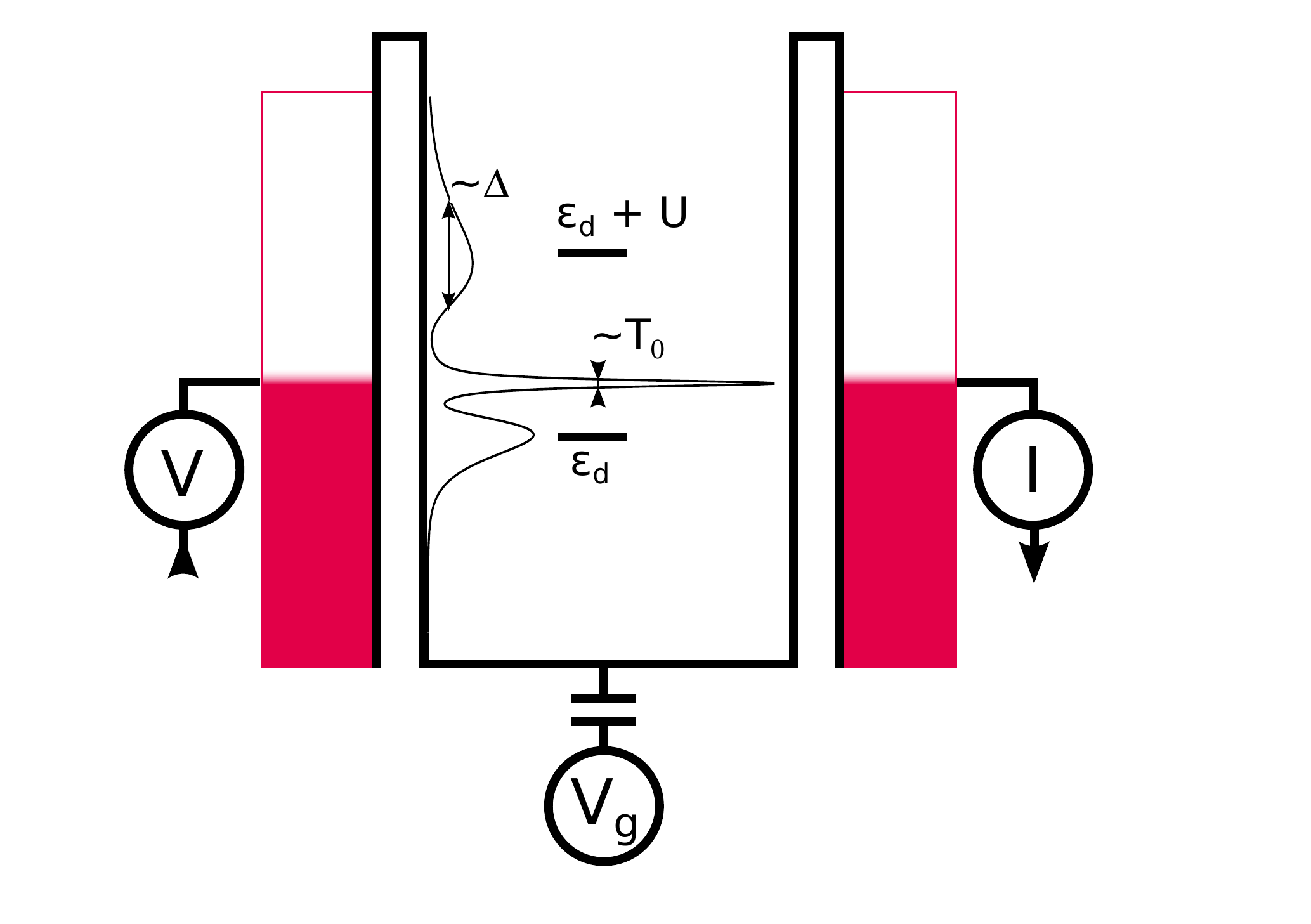}
\caption
{
  {\em (Color online)}
A strongly correlated quantum dot with charging energy $U\gg \Delta$ and level energy $\varepsilon_{d}$ connected 
to leads via tunnel barriers.
The gate voltage $V_{g}\sim\varepsilon_{d}$ allows changing occupation of the dot $n_{d}$ from
$n_{d}=1$ for $\varepsilon_{d}=-U/2$ to $n_{d}=0$ through a mixed valence regime with $n_{d}\approx 0.5$ 
for $\varepsilon_{d}\approx 0$. In the singly occupied configuration, shown here for $\varepsilon_{d}\approx -U/2$, 
the dot has a well defined spin $1/2$ and the Coulomb blockade excitations at 
$\varepsilon_{d}$ and $\varepsilon_{d}+U$ correspond to removing or adding an electron.
The coupling of the spin $1/2$ to the leads results in the Kondo effect, which is manifested
by the appearance of an additional many-body Kondo resonance at the Fermi level $\epsilon_{F}=0$ at low 
temperatures $T\leq T_{0}$ . This resonance is also reflected as a zero bias anomaly
in the non-linear conductance $dI/dV$ in experiments. \cite{Goldhaber1998a} 
  \label{cartoon-quantum-dot}
}
\end{figure}

We consider the simplest model of a correlated quantum dot, the single-level Anderson model given by the 
Hamiltonian
\begin{eqnarray}
H &=& \sum_{\sigma}\varepsilon_{d\sigma}n_{d\sigma} -g\mu_{\rm B}B\,s_{z}^{d} + U n_{d\uparrow}n_{d\downarrow}
\nonumber\\
&+& \sum_{k\alpha\sigma}\epsilon_{k\alpha}c_{k\alpha\sigma}^{\dagger}c_{k\alpha\sigma}+ 
\sum_{k\alpha\sigma}(t_{\alpha}c_{k\alpha\sigma}^{\dagger}d_{\sigma}+ h.c.).\label{model}
\end{eqnarray}
Here, $\varepsilon_{d}$ is the level energy, related to the gate voltage $V_{g}$ in the quantum dot via $\varepsilon_{d}\sim eV_{g}$ 
(see Fig.~\ref{cartoon-quantum-dot}), $B$ is a local magnetic field acting 
on the quantum dot with $s_{z}^{d}=\frac{1}{2}(n_{d\uparrow}-n_{d\downarrow})$,
$U>0$ is the Coulomb charging energy, $\sigma$  labels the spin, and  
$\alpha=L,R$ labels left and right electron lead states with kinetic 
energies $\epsilon_{k\alpha}$. The couplings of the dot to the leads  are
denoted by $\Delta_{\alpha}(\omega)=\pi \rho_{\alpha}(\omega)|t_{\alpha}|^{2}$, 
where $\rho_{\alpha}(\omega)=\sum_{k}\delta(\omega-\epsilon_{k\alpha})$ is
the density of states of lead $\alpha$. For simplicity we assume a constant density of states $\rho_{\alpha}=N_{\rm F}=1/2D$ 
with half-bandwidth $D=1$ so that $\Delta_{\alpha}=\pi N_{\rm F}t_{\rm \alpha}^{2}$. By using even and 
odd parity combinations of left and right lead states, model (\ref{model}) is reduced to a single-channel Anderson model 
with a resonant level half-width at half-maximum given by $\Delta=\Delta_{\rm L}+\Delta_{\rm R}$. The spectral function of the latter
model is required in the calculation of the linear conductance, which we describe next.

\section{NRG calculation of conductance}
\label{sec:nrg-cond}

\begin{figure}
\includegraphics[width=\linewidth,clip]{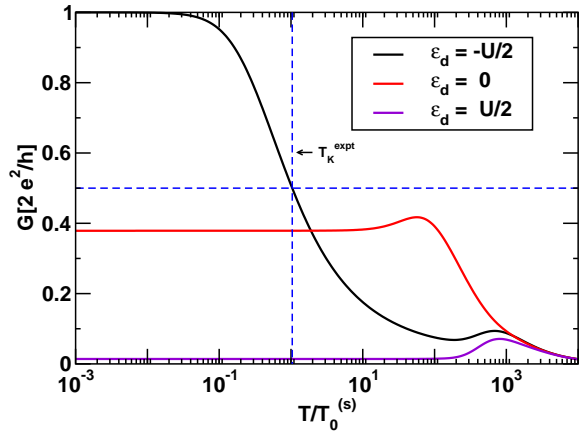}
\caption
{
  {\em (Color online)}
  Linear conductance $G(T)$ versus $T/T_{\rm 0}^{s}$ for $U/\Delta=16$ and several values of $\varepsilon_{d}=-U/2, 0, +U/2$ using
  the approach of Ref.~\onlinecite{Yoshida2009}, with $T_{0}^{s}$ defined by Eqs.(\ref{eq:chi0}-\ref{eq:t0s-large-U}).
  We also indicate with horizonatl and vertical dashed lines the extraction of the experimental Kondo scale, $T_{\rm K}^{\rm expt}$, 
  from the mid-valley Kondo conductance (i.e. that for $\varepsilon_{d}=-U/2$) via $G(T=T_{\rm K}^{\rm expt})=G(0)/2$.
  NRG parameters were for $\Lambda = 4$, with an  energy cut-off $e_{c}(\Lambda=4)=30$ and  $n_{z}=2$.
  \label{fig-G-versus-T}
}
\end{figure}
The linear response electrical conductance $G(T,B)$ of (\ref{model}) is given by \cite{Hershfield1991,Meir1993}
\begin{eqnarray}
G(T,B)&=& \frac{e^{2}}{h}\int d\omega\; 
\left(-\frac{\partial f}{\partial \omega}\right) \;\sum_{\sigma}{\cal T}_{\sigma}(\omega,T,B),\label{linear-cond}
\end{eqnarray}
where 
\begin{equation}
{\cal{T}}_{\sigma}(\omega,T,B)=4\pi \frac{\Delta_{L}\Delta_{R}}{\Delta_{L}+\Delta_{R}}A_{\sigma}(\omega,T,B)
\end{equation}
is the transmission function for spin $\sigma$ electrons. It can be calculated from the single-particle spectral 
function of the dot 
$A_{\sigma}(\omega,T,B)
= -{\rm Im}[G_{d\sigma}(\omega+i\delta)]/\pi,
$
where 
$G_{d\sigma}(\omega+i\delta)=\langle\langle d_{\sigma}; 
d_{\sigma}^{\dagger}\rangle\rangle$
is the Fourier transform of the retarded single-particle Green function of (\ref{model}). In Eq.~(\ref{linear-cond}),
$e$ and $h$ are the electronic charge and Planck's constant respectively, and $f(\omega)=[1+\exp(\beta\omega)]^{-1}$ 
is the Fermi function. 

We use the NRG to evaluate the spectral function $A_{\sigma}(\omega,T,B)$ via the Lehmann representation
\begin{align}
A_{\sigma}(\omega,T,B) &= \frac{1}{Z}\sum_{m,n}|M_{mn}^{\sigma}|^{2}(e^{-\beta E_{m}} +e^{-\beta E_{n}})\nonumber\\
                      &\times\delta(\omega -(E_{m}-E_{n})),\label{lehmann}
\end{align}
where $M_{mn}^{\sigma}$ are the matrix elements of the spin $\sigma$ local d-electron operator between eigenstates $|m\rangle$ and $|n\rangle$ with
energies $E_{m}$ and $E_{n}$ and $Z=\sum_{m}\exp(-\beta E_{m})$ is the partition function
(see Ref.~\onlinecite{Bulla2008} for details). The usual approach is to broaden the discrete spectral function in Eq.~(\ref{lehmann})
with Gaussians or Logarithmic Gaussians in order to obtain a smooth function,\cite{Sakai1989,Bulla2001,Bulla2008} 
which is then substituted into Eq.~(\ref{linear-cond}),
thereby yielding $G(T,B)$ after a numerical integration. A more accurate procedure, introduced in Ref.~\onlinecite{Yoshida2009},
is to substitute the discrete representation in Eq.~(\ref{lehmann}) directly into Eq.~(\ref{linear-cond}) resulting in the expression 
\begin{equation}
G(T,B) = \frac{\gamma\beta}{Z}\sum_{\sigma}\sum_{m,n}|M_{mn}^{\sigma}|^{2}\frac{1}{e^{\beta E_{m}} +e^{\beta E_{n}}},\label{yoshida-cond}
\end{equation}
where $\gamma=4\pi \frac{e^2}{h}\frac{\Delta_{L}\Delta_{R}}{\Delta_{L}+\Delta_{R}}$. This avoids errors from numerical integrations and
from an artificial broadening of the spectral function and has been shown to give accurate results for the conductance. \cite{Yoshida2009}
A similar procedure has been used in Ref.~\onlinecite{Weichselbaum2007} to extract $c_{T}$ for the symmetric Anderson model 
to within $5\%$ accuracy. This uses a full density matrix evaluation of the spectral function\cite{Weichselbaum2007,Peters2006} 
within the complete basis set\cite{Anders2005} and is computationally more intensive than the approach which we use here, 
whose computational complexity is comparable to that of evaluating a local static correlation function.

For the remainder of this paper we shall set the $g$ factor $g$, Bohr magneton $\mu_{B}$, Planck's constant $h$, electric charge $e$, and 
Boltzmanns constant $k_{\rm B}$ to unity, and also assume symmetric coupling to the leads 
($\Delta_{L}=\Delta_{R}=\Delta/2,\gamma=\pi\Delta$). A finite asymmetry $\Delta_{L}\neq \Delta_{R}$ only influences the value of $G(0,0)$, but
not our results for $c_{B}$ and $c_{T}$. Note that in experiments on quantum dots,\cite{Kretinin2011,Goldhaber1998} 
the extracted full width at half maximum of the Coulomb blockade peaks is given by $\Gamma = 2\Delta$.

In evaluating Eq.~(\ref{yoshida-cond}), we used $z$-averaging \cite{Oliveira1994} 
within the band discretization scheme of Ref.~\onlinecite{Campo2005} 
(see Appendix~\ref{appendix:campo}). A discretization parameter of $\Lambda=4$ with $n_{z}=2$ values for the $z$-averaging
was used and the cut-off for the rescaled energies at each NRG iteration was set to $e_{c}(\Lambda=4)=30$. Figure~\ref{fig-G-versus-T} shows
typical examples for $G(T)$ versus $T/T_{0}$ at $B=0$ for a strongly correlated quantum dot ($U/\Delta=16 \gg 1$) in
the Kondo ($\varepsilon_{d}=-U/2$), mixed valence ($\varepsilon_{d}=0$) and empty orbital ($\varepsilon_{d}=U\gg \Delta$)
regimes. The scale $T_{0}$ is defined from the $T=0$ susceptibility of the Anderson model (\ref{model})
\begin{equation}
\chi(0)=1/4T_{0}\label{eq:chi0}
\end{equation}
for all $U$ and $\varepsilon_{d}$. For the case of particle-hole symmetry ($\varepsilon_{d}=-U/2$) and strong correlations $U\gg \pi\Delta$, one also has from the Bethe ansatz solution for $\chi(0)$ an analytic expression for $T_{0}$
\begin{equation}
T_{\rm 0}(\varepsilon_{d}=-U/2)\equiv T_{0}^{(s)}\approx\sqrt{U\Delta/2}e^{-\pi U/8\Delta + \pi \Delta/2U},\label{eq:t0s-large-U}
\end{equation}
within corrections which are exponentially small in 
$U/\pi\Delta$ (see Ref.~\onlinecite{Hewson1997}).

\begin{table}
\begin{ruledtabular}
\begin{tabular}{llll}
Fitting range & $R^2$   & $c_T$   & \% error\\
\colrule
$10^{-5}T_{0}\le T\le T_0$      &  0.818      &  0.7447  &   87     \\
$10^{-5}T_{0}\le T\le 0.1 T_0$   &  0.9969     &  5.1277  &   16     \\
$10^{-5}T_{0}\le T\le 0.05 T_0$  &  0.99965    &  5.8002  &   4.7   \\
$10^{-5}T_{0}\le T\le 0.02 T_0$  &  0.999980   &  6.0820  &   0.086 \\
$10^{-5}T_{0}\le T\le 0.01 T_0$  &  0.9999894  &  6.1459  &   0.96  \\
$10^{-5}T_{0}\le T\le 0.005 T_0$ &  0.99985    &  6.1468  &   0.98  \\
$10^{-5}T_{0}\le T\le 0.001 T_0$ &  0.9381     &  6.2034  &   1.91
\end{tabular}
\end{ruledtabular}
\caption{Optimal temperature range for fitting the
conductance $G(T,0)$ to the Fermi liquid form $f(T/T_{0})=a(1 - c_{T}(T/T_{0})^2)$ for $U/\Delta=12$ and
$\varepsilon_{d}=-U/2$ using a goodness of fit based on the value of $R^2$ . The latter
is defined by
$R^2 = 1 - \frac{\sum_{i=1}^{n} (y_i-f(x_i))^2}{\sum_{i=1}^{n} (y_i - <y>)^2 }$, where $x_{i}=T_{i}/T_{0}$, $y_{i}=G(T_{i},0)$, 
$\langle y\rangle=\frac{1}{n}\sum_{i=1}^{n}y_{i}$ and the number of data points in the fitting ranges was 
$n\approx 200$. The value $R^{2}=1$ would correspond to a perfect fit to the Fermi liquid form. 
The \% error in $c_{T}$ in the last column is defined 
by    $\text{\% error} = 100\cdot|\frac{c_T-c_{T,\text{exact}}}{c_{T,\text{exact}}}|$
where $c_{T,\text{exact}}$ is the exact value at particle-hole symmetry given by Eq.~(\ref{eq:ct-symmetric-general-u}). 
From this table we see that the optimal range which maximizes $R^2$ and the accuracy of $c_{T}$ is close to $T\le 0.02T_{0}$.
The NRG calculations used $\Lambda=4$, $n_{z}=2$ and an energy cut-off $e_{c}(\Lambda)=30$.
}\label{table-one}
\end{table}
The Kondo scale $T_{0}$ is useful in analytic calculations of $c_{T}$ and $c_{B}$ about the Fermi liquid fixed point at $T=0$, such as
those in Sec.~\ref{sec:fermi liquid}. With this definition, the meaning of the coefficients $c_{T}$ and $c_{B}$ is 
fixed by
\begin{align}
\frac{G(T,B=0)}{G(0,0)}=& 1-{c}_{\rm T}\left(\frac{T}{T_{\rm 0}}\right)^{2}, & (T\ll T_{0}),\label{eq:fl-region-ct}\\
\frac{G(T=0,B)}{G(0,0)}=& 1-{c}_{\rm B}\left(\frac{B}{T_{\rm 0}}\right)^{2}, &(B\ll T_{0}).\label{eq:fl-region-cb}
\end{align}
In extracting $c_{B}$ we do not use Eq.~(\ref{eq:fl-region-cb}), but instead use the Fermi liquid result in 
Eq.~(\ref{eq:cb_fermi_liquid}) of Sec.~\ref{sec:fermi liquid}. This allows $c_{B}$ to be obtained directly 
from the $T=0$ occupancy of the $d$-level, a quantity that can be calculated
to high accuracy within the NRG. The coefficient $c_{T}$ is extracted numerically by fitting $G(T,0)$ in 
the range $10^{-5}T_{0}\le T\le 2\times 10^{-2}T_{0}$ to the Fermi liquid 
form in Eq.~(\ref{eq:fl-region-ct}) with $T_{0}$ as defined in Eq.~(\ref{eq:chi0}). The range $T\le 0.02T_{0}$ was found
optimal for this purpose, as we now describe. \footnote{The lower bound of this range $T_{\rm min}=10^{-5}T_{0}$ is the lowest 
temperature for NRG calculations. Provided it is small enough, its precise value does not affect the quality of the fits significantly.}
Specifically, we fit the NRG results for $G(T,0)$ in the above range to $f(x)=a(1-c_{T}x^{2})$ 
where $x=T/T_{0}$. We find that $a=2\pm 10^{-5}$ at the particle-hole symmetric point, with $a=2$ (in units of $e^2/h$) 
being the exact result from the Friedel sum rule. The effect of the fitting range on the accuracy of the extracted $c_{T}$ and the degree of
confidence in the fermi liquid form $f(x)=a(1-c_{T}x^{2})$ may be ascertained quantitatively by calculating the $R$ squared coefficient
$R^{2}$ (also called the coefficient of determination and defined in Table~\ref{table-one}). 
Table~\ref{table-one} lists $R^2$, together with the extracted $c_{T}$ and the $\%$ error in 
$c_{T}$ for different fitting ranges. We see that the range $10^{-5}T_{0}\le T\le 0.02T_{0}$ is close to maximizing both $R^{2}$ and the accuracy of $c_{T}$ 
[as compared to the exact result in Eq.~(\ref{eq:ct-symmetric-general-u}) of Sec.~{\ref{sec:fermi liquid}}]. We therefore used this 
range throughout, also for the asymmetric cases. Care is needed in the choice of the cut-off $e_{c}(\Lambda)$ in order to obtain correct results for $G(T)$ 
in the low-temperature limit when using Eq.~(\ref{yoshida-cond}). If $e_{c}(\Lambda)$ is chosen to be too small, 
the correct saturation behavior of $G(T,0)$ in the low-temperature limit (i.e. the ``leveling-off'' of the conductance) is not
obtained. In this case, a fit of $G(T,0)$ to $f(T/T_{0})$ shows a drop in $R^{2}$ to small values, indicating a problem. 
This is remedied by increasing $e_{c}(\Lambda)$ (the used value $e_{c}(\Lambda=4)=30$ was sufficient, whereas $e_{c}(\Lambda=4)=12$, for example,
is not). Thus, the $R^2$ criterion can be a useful check on appropriate choices of cut-off when evaluating the 
conductance via Eq.~(\ref{yoshida-cond}).

In experiments on Kondo correlated quantum dots, $T_{0}$ is not measurable, and instead one extracts a Kondo scale, $T_{\rm K}^{\rm expt}$,  
from the temperature dependence of the $B=0$ conductance via
\begin{equation}
G(T=T_{\rm K}^{\rm expt})=G(0)/2.
\end{equation}
In principle, this $T_{\rm K}^{\rm expt}$ can be extracted for each gate voltage (i.e., for each $\varepsilon_{d}$), but in practice,
it is usually extracted only at mid-valley ($\varepsilon_{d}=-U/2$) 
where one is sure to be in the Kondo regime for large $U/\Delta$. This is illustrated in 
Fig.~\ref{fig-G-versus-T} by the dashed lines.

With this definition of $T_{\rm K}^{\rm expt}$, one extracts the experimentally measured coefficients $c_{T}^{\rm expt}$ and $c_{B}^{\rm expt}$ via
\begin{align}
\frac{G(T,B=0)}{G(0,0)}=&1-{c}_{\rm T}^{\rm expt}\left(\frac{T}{T_{\rm K}^{\rm expt}}\right)^{2}, &(T\ll T_{\rm K}^{\rm expt}),\\
\frac{G(T=0,B)}{G(0,0)}=&1-{c}_{\rm B}^{\rm expt}\left(\frac{B}{T_{\rm K}^{\rm expt}}\right)^{2}, &(B\ll T_{\rm K}^{\rm expt}).
\end{align}
For the particle-hole symmetric Anderson model, the coefficients $c_{T}^{\rm expt}$ and $c_{B}^{\rm expt}$ are related to $c_{T}$ and $c_{B}$ via
\begin{eqnarray}
c_{T}^{\rm expt} & = c_{T}\left(\frac{T_{\rm K}^{\rm expt}}{T_{\rm 0}^{(s)}}\right)^{2}\label{exp-ct},\\
c_{B}^{\rm expt} & = c_{B}\left(\frac{T_{\rm K}^{\rm expt}}{T_{\rm 0}^{(s)}}\right)^{2}\label{exp-cb}.
\end{eqnarray}
For a precise translation of theoretical calculations of $c_{B}$ and $c_{T}$ in terms of $T_{0}$, into 
experimentally measured ones in terms of $T_{\rm K}^{\rm expt}$, one therefore requires the ratio $T_{\rm K}^{\rm expt}/T_{0}^{s}$ 
at mid-valley for all charging energies ($U/\Delta$), which we supply in Sec.~\ref{sec:results}.

\section{Fermi liquid results for $G(T,B)$}
\label{sec:fermi liquid}
For the case of particle-hole symmetry, the coefficient $c_{T}$ is known for arbitrary $U/\Delta$ 
within renormalized perturbation theory about the Fermi liquid fixed point. \cite{Hewson2006} 
The expression is given by
\begin{equation}
c_{T}=\frac{\pi^4}{12}\frac{1+2(R-1)^2}{R^2},\label{eq:ct-symmetric-general-u}
\end{equation}
where $R$ is the Wilson ratio [defined in Eq.~(\ref{eqRR4})]. In the limit of strong correlations,
$U/\Delta \gg 1$, the Wilson ratio approaches $2$ and $c_{T}$ takes the well known universal Kondo 
value $c_{T}=\pi^{4}/16$ (see Ref.~\onlinecite{Nozieres1974,Costi1994}). In the opposite limit 
$U/\Delta\rightarrow 0$ the Wilson ratio tends to $1$ and $c_{T}$ acquires the value $\pi^{4}/12$. 
Evaluation of Eq.~(\ref{eq:ct-symmetric-general-u}) for general $U/\Delta$ requires knowledge of 
$R$, either from Bethe ansatz or from NRG.

Fermi liquid theory allows an exact analytic expression for $c_{B}$ to be obtained for all $U$ and 
$\varepsilon_{d}$.  For this purpose we use the Friedel sum rule $A_{\sigma}(0,B)=\sin^{2}(\pi n_{d\sigma}(B))/\pi\Delta$,
where $n_{d\sigma}(B)$ is the spin $\sigma$ local level occupancy in a small finite magnetic field $B\ll T_{0}$ at $T=0$.
Using $n_{d\sigma}(B)=n_{d}/2 + \sigma\alpha B$, where $n_{d}$ is the total occupancy at $B=0$, and the fact that
$\alpha = \frac{1}{2}\frac{n_{d\uparrow}(B)-n_{d\downarrow}(B)}{B}=\chi(0)$ we easily find the exact result (correct
to order $B^{2}$)
\begin{eqnarray}
\frac{G(0,B)}{G(0,0)} &=& 1 - \pi^{2}\chi^{2}(0)B^{2}(1-\cot^{2}(\frac{\pi n_{d}}{2}))\\
&=& 1- c_{B}\left(\frac{B}{T_{0}}\right)^{2},\nonumber\\
c_{B} & = & \frac{\pi^2}{16}(1-\cot^{2}(\frac{\pi n_{d}}{2})),\label{eq:cb_fermi_liquid}
\end{eqnarray}
where $\chi(0)=1/4T_{0}$ has been used. Note that at the particle-hole
symmetric point ($\varepsilon_{d}=-U/2$), where $n_{d}=1$,
$c_{B}$ takes the universal Kondo value $\pi^{2}/16$ for all $U$. \cite{Konik2002}
This universal result [obtained also within SPT, see Eq.~(\ref{cb-expression})] 
could  be tested in semiconductor quantum dots that can be tuned through complete valleys.
It would then acquire the value $\pi^{2}/16$ at mid-valley for any valley.
The expression in Eq.~(\ref{eq:cb_fermi_liquid}) also shows that $c_{B}$ decreases 
monotonically with increasing gate voltage away from mid-valley, with $c_{B}$ becoming 
negative on entering the mixed valence regime (which we define by the average occupation being $n_{d}=0.5$).

\section{SPT calculation}
\label{sec:SPT}
The SPT approach \cite{Munoz2013} is based on a renormalized perturbation 
theory on the Keldysh contour \cite{Hewson1993,Oguri2005} using dual fermions \cite{Rubtsov2008,Hafermann2009,Jung2012}. This approach can be shown to be
current conserving by construction \cite{Munoz2013} even in the nonlinear
response regime, as opposed to finite-order perturbation theory
in the bare parameters \cite{Hershfield1992}. 
We compare the results of this theory for the linear conductance with NRG calculations. The reference system is
the interacting particle-hole symmetric Anderson model characterized by the renormalized Coulomb interaction $\tilde{u}=z \Gamma_{0}/\pi\Delta$,
where $z$ is the wave function renormalization constant, and $\Gamma_{0}(U)\equiv\Gamma_{\uparrow,\downarrow;\downarrow,\uparrow}(0,0;0,0)$ the four-point
vertex. In order to obtain results for the asymmetric model, an 
expansion in $\tilde{\varepsilon}_{d}\equiv z(\varepsilon_{d}+U/2)/z\Delta = (\varepsilon_{d}+U/2)/\Delta$ is carried 
out for the local level Green function up to a given order in $\tilde{\varepsilon}_{d}$ and $\tilde{u}$, currently up to
order $\tilde{u}^{2}\tilde{\varepsilon}_{d}^{2}$. As $z\rightarrow 0$ with increasing $U/\Delta$ such that in the symmetric 
case $\tilde{u}\rightarrow 1$, it follows that $\tilde{u}^2\tilde{\epsilon}_d^2$ increases with growing  $U/\Delta$  at  fixed $\epsilon_d/\Delta$.
An outline of the method is presented in appendix B, with full details available in the Supplemental Material of Ref.\onlinecite{Munoz2013}. 
The expression for $c_{T}$, given in Ref.~\onlinecite{Munoz2013}, and that
for $c_{B}$, derived in Appendix~\ref{appendix:cb calculation}, are given by
\begin{eqnarray}
c'_{T} &=& \frac{\pi^{4}}{12}\frac{1+2\tilde{u}^2 +(1-\tilde{u})(5\tilde{u}-3)\tilde{\varepsilon}_{d}^{2}}
{(1+\tilde{u})^{2}(1+(1-\tilde{u})^2\tilde{\varepsilon}_{d}^{2})^2},\label{ct-expression}\\
c'_{B} &=& \frac{\pi^{2}}{16}\frac{1-3(1-\tilde{u})^{2}\tilde{\varepsilon}_{d}^{2}}{(1+(1-\tilde{u})^2\tilde{\varepsilon}_{d}^{2})^2},\label{cb-expression}
\end{eqnarray}
where the apostrophe on these indicates that they are evaluated by using the susceptibility Kondo scale of the reference system (symmetric Anderson
model), i.e., $c'_{T}, c'_{B}$ are defined via
\begin{align}
\frac{G(T,B=0)}{G(0,0)} = &1 - c'_{T}\left(\frac{T}{T_{0}^{(s)}}\right)^{2}, &(T\ll T_{0}^{s}),\label{gt-spt}\\
\frac{G(T=0,B)}{G(0,0)} = &1 - c'_{B}\left(\frac{B}{T_{0}^{(s)}}\right)^{2}, &(B\ll T_{0}^{s}),\label{gb-spt}
\end{align}
where $T_{0}^{(s)}=1/4\chi(0)$ is the susceptibility Kondo scale for the symmetric model and is given explicitly within SPT by Eq.~{\ref{eq:spt-t0}} in
Sec.~{\ref{appendix:cb calculation}}. In order to compare the above results with those from NRG, we need to relate the 
renormalized Coulomb interaction, $\tilde{u}$, appearing in the former, to the bare Coulomb interaction, 
$U$, appearing in the latter. As outlined in Appendix~\ref{appendix:cb calculation}, from Eq.~(\ref{eqRR5}) 
this relation is given by $\tilde{u}=R-1$ for the symmetric Anderson model, 
where the Wilson ratio, $R$, is calculated for given $U/\Delta$ from the exact Bethe ansatz expressions for the
susceptibility and specific heat\cite{Zlatic1983} of the fully interacting symmetric Anderson model \cite{Tsvelick1983}.
Notice that upon substituting $\tilde{u}=R-1$ into the SPT expression Eq.~\ref{ct-expression}, 
for the particle-hole symmetric limit $\tilde{\varepsilon}_{d}=0$, it
reduces to the exact Fermi liquid result Eq.~\ref{eq:ct-symmetric-general-u}.

\section{Results}
\label{sec:results}
\subsection{Symmetric case ($\varepsilon_{d}=-U/2$)}
\label{subsec:symmetric case}
\begin{figure}
\includegraphics[width=\linewidth,clip]{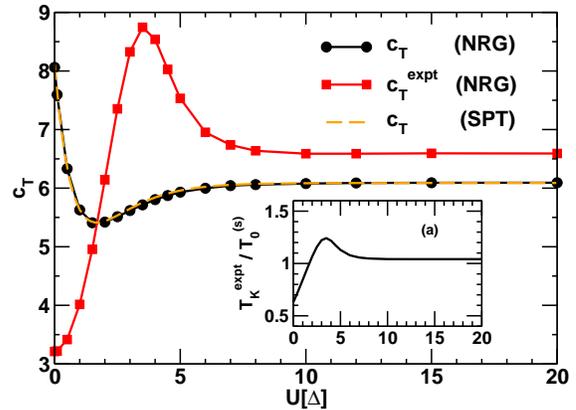}
\caption
{
  {\em (Color online)}
  $c_{T}$ vs $U/\Delta$ for the symmetric Anderson model calculated within NRG (solid lines with symbols) and SPT (dashed line). Filled circles 
  show $c_{T}$ using the susceptibility scale $T_{0}^{(s)}$, while filled squares show $c_{T}$ upon using the scale from the conductance $T_{\rm K}^{\rm expt}$.
  By comparing the value of $c_{T}$ at $U/\Delta \gg 1$ with the exact one $c_{T}=\pi^4/16 \approx 6.088$, we
  estimate the relative error in the NRG calculation of $c_{T}$ to lie below $0.2\%$, considerably more accurate than previous estimates.
  \cite{Costi1994,Weichselbaum2007}
  Inset (a): ratio $T_{\rm K}^{\rm expt}/T_{0}^{(s)}$ vs $U/\Delta$. For $U/\Delta \gg 1$, the ratio  $T_{\rm K}^{\rm expt}/T_{0}^{(s)}$ approaches $1.04$.
  NRG parameters were for $\Lambda = 4$ with an  energy cut-off $e_{c}(\Lambda=4)=30$ and $n_{z}=2$.
  \label{fig-symmetric-compare}
}
\end{figure}

\begin{figure}
\includegraphics[width=\linewidth,clip]{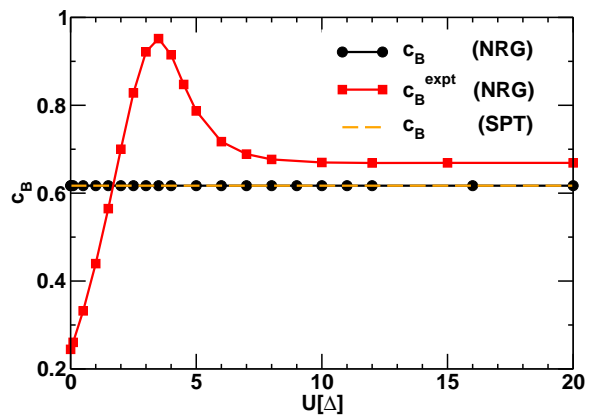}
\caption
{
  {\em (Color online)}
  $c_{B}$ (filled circles) and $c_{B}^{\rm expt}$ (filled squares) vs $U/\Delta$ for the symmetric Anderson model calculated within NRG 
  NRG parameters were for $\Lambda = 4$ with an  energy cut-off $e_{c}(\Lambda=4)=30$ and  $n_{z}=2$. SPT (dashed curve) also recovers
  the value $c_{B}=\pi^{2}/16$ for particle-hole symmetry.
  \label{fig-symmetric-compare-cb}
}
\end{figure}

Figure~\ref{fig-symmetric-compare} shows $c_T$ vs $U$ for the symmetric model, both in terms of the scale $T_{0}(\varepsilon_{d}=-U/2)\equiv T_{0}^{(s)}$
and in terms of the Kondo scale from the conductance $T_{\rm K}^{\rm expt}$ (i.e., $c_{T}^{\rm expt}$, discussed below). The former is compared with the corresponding 
SPT prediction in Eq.~(\ref{ct-expression}) and we see very good agreement between this and the NRG calculations for all $U/\Delta$. 
For $U/\Delta \gtrsim 6$, the value of $c_{T}$ remains within $2\%$ of the
the universal Kondo value $c_{T}=\pi^{4}/16=6.088$. The value $U/\Delta=\pi \approx 3$ separates the weakly correlated 
($U/\pi\Delta < 1$) from the strongly correlated regime ($U/\pi\Delta>1$). \cite{Hewson1997} We see that
in the moderately correlated regime $6\gtrsim U/\Delta\gtrsim 1$, the deviation $c_{T}$ from the Kondo value increases, 
eventually reaching $7\%$ at $U/\Delta \approx 3$. 
For weakly correlated (non-Kondo) quantum dots with $U/\pi\Delta \lesssim 1$, $c_{T}$ first decreases with decreasing $U$, 
reaches a minimum at $U/\Delta\approx 1.7$, and then increases to its non-interacting value
of $\pi^{4}/12 \approx 8.117$ at $U=0$. Note that this latter value differs by more than $30\%$ from the Kondo value at $U\gg \Delta$. 

In Fig.~\ref{fig-symmetric-compare}(a) we show the ratio  $T_{\rm K}^{\rm expt}/T_{0}^{(s)}$ vs $U$. This ratio allows obtaining 
$c^{\rm expt}_{T}$ from Eq.~(\ref{exp-ct}), the coefficient measured in experiments, and which we 
show in  Fig.~\ref{fig-symmetric-compare}.
Note the very different behavior between $c_{T}^{\rm expt}$ and $c_{T}$ for charging energies $U/\Delta \lesssim 5$. This is due to the strong
dependence of $T_{\rm K}^{\rm expt}/T_{0}^{(s)}$ on $U/\Delta$ in this range of charging energies. In particlular $c_{T}^{\rm expt}$ acquires a 
maximum value of $\approx 8.75$ at $U/\Delta\approx 3.5$.
Since $T_{\rm K}^{\rm expt}/T_{0}^{(s)}\approx 1.041$ for $U/\Delta\gtrsim 10$ [Fig.~\ref{fig-symmetric-compare}(a)], $c_{T}^{\rm expt}\approx 6.58$ 
for strongly correlated quantum dots in the Kondo regime.\footnote{A. Weichselbaum and M. Hanl find a similar 
value $T_{\rm K}^{\rm expt}/T_{0}^{(s)}\approx 1.06\pm 0.03$ within a full density matrix approach to the spectral function\cite{Weichselbaum2007} 
(private communication). The present result improves on an earlier estimate for the $S=1/2$ Kondo model \cite{Micklitz2006} which yielded 
$T_{\rm K}^{\rm expt}/T_{0}^{(s)}\approx 0.94$.} In contrast, for $U/\Delta \lesssim 10$, the scale $T_{\rm K}^{\rm expt}$ 
differs appreciably from $T_{0}^{(s)}$. Thus, even for nominally Kondo correlated 
quantum dots with $U/\Delta \approx 4.5$, such as those in Ref.~\onlinecite{Kretinin2011}, one finds from Fig.~\ref{fig-symmetric-compare}(a)
that $T_{\rm K}^{\rm expt}/T_{0}^{(s)}\approx 1.18$, so one should expect $c_{T}^{\rm expt}\approx 7.5$, which is somewhat 
larger than the extracted value $c_{T}^{\rm expt}\approx 5.6\pm 1.2$.\cite{Kretinin2011} 

As discussed in Sec.~\ref{sec:fermi liquid}, $c_{B}$ is independent of the charging energy $U$ for the particle-hole symmetric case, 
where it takes the value $\pi^{2}/16\approx 0.617$, which is also recovered exactly within SPT [see Eq.~(\ref{cb-expression})].
However, experiments use the scale $T_{\rm K}^{\rm expt}$ and measure $c_{B}^{\rm expt}$ as  
given by Eq.~(\ref{exp-cb}). This depends on $U$ through the ratio
of Kondo scales $T_{\rm K}^{\rm expt}/T_{0}^{(s)}$. For completeness, we therefore show the $U$ dependence of  $c_{B}^{\rm expt}$ 
in Fig~\ref{fig-symmetric-compare-cb}. 
For $U/\Delta\approx 4.5$, relevant for the experiments in Ref.~\onlinecite{Kretinin2011}, 
we find $c_{B}^{\rm expt}\approx 0.89$ significantly smaller 
than the value $c_{B}^{\rm expt}\approx 5.1$ extracted from the measurements. As discussed in that paper, the large discrepancy between the 
measured and predicted  values of $c_{B}^{\rm expt}$ could indicate the importance of the large spin-orbit interaction present in the
InAs quantum dots investigated in Ref.~\onlinecite{Kretinin2011}.

\subsection{Asymmetric case ($\varepsilon_{d}>-U/2$)}
\label{subsec:asymmetric case}

\subsubsection{NRG results}
\label{subsubsec:asymmetric NRG}

\begin{figure}
\includegraphics[width=\linewidth,clip]{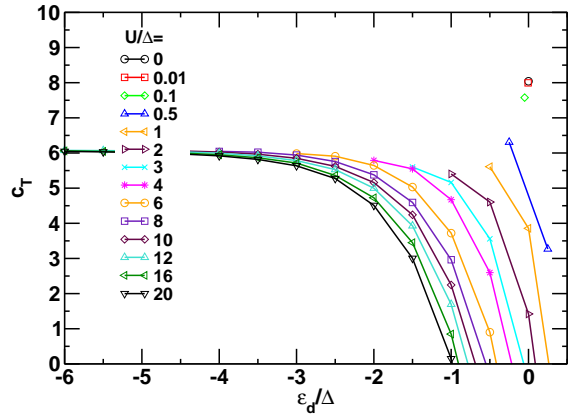}
\caption
{
  {\em (Color online)}
  $c_{T}$ vs $\varepsilon_{d}$ (in intervals of $0.5\Delta$) with $\varepsilon_{d}\ge -U/2$ for several $U/\Delta$, ranging from strong, $U\gg\Delta$, 
  to weak, $U\ll\Delta$, correlations, and using the scale $T_{0}$. 
  NRG parameters were for $\Lambda = 4$ with an  energy cut-off $e_{c}(\Lambda=4)=30$ and  $n_{z}=2$.
  \label{fig-ct-all-susc}
}
\end{figure}
\begin{figure}
\includegraphics[width=\linewidth,clip]{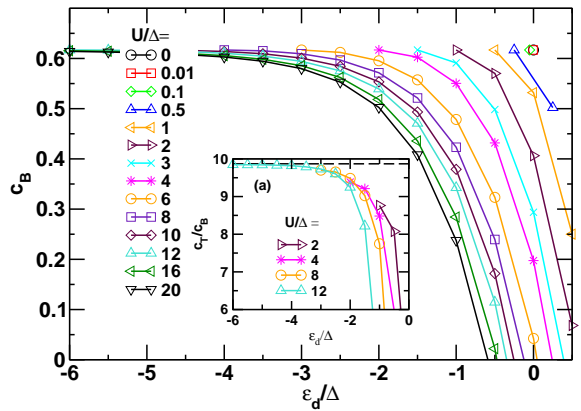}
\caption
{
  {\em (Color online)}
  $c_{B}$ vs $\varepsilon_{d}$ (in intervals of $0.5\Delta$) with $\varepsilon_{d}\ge -U/2$ for several $U/\Delta$, ranging from strong, $U\gg\Delta$, 
  to weak, $U\ll\Delta$, correlations, and using the scale $T_{0}$. 
  NRG parameters were for $\Lambda = 4$ with an  energy cut-off $e_{c}(\Lambda=4)=30$ and  $n_{z}=2$.
  Inset (a) shows $c_{T}/c_{B}$ vs $\varepsilon_{d}/\Delta$ for selected $U/\Delta$. The dashed line is a guide to the eye and 
  represents the universal Kondo value $c_{T}/c_{B}=\pi^2$ reached in the limit $U/\Delta\rightarrow\infty$ and particle-hole symmetry.
  \label{fig-cb-all-susc}
}
\end{figure}
For completeness, we show the dependence of $c_T$ on $\varepsilon_{d}$ for $\varepsilon_{d}\geq -U/2$ for $U$ ranging from weakly ($U/\Delta\ll 1$) to strongly 
($U/\Delta\gg 1$) correlated quantum dots in Figure~\ref{fig-ct-all-susc}. 
For strong correlations $U/\Delta \gg 1$, $c_{T}$ decreases monotonically 
with increasing deviations from the Kondo regime, eventually becoming negative after the mixed valence regime is reached. 
A similar behavior is seen in the local level dependence of $c_{B}$, shown in Fig.~\ref{fig-cb-all-susc}. 
In Fig.~\ref{fig-cb-all-susc}(a) we show the ratio $c_{T}/c_{B}$ versus $\varepsilon_{d}/\Delta$ for selected $U/\Delta$ which approaches
the value $\pi^2$ at particle-hole symmetry and $U/\Delta\gg 1$. Notice that for correlated quantum dots in the Kondo regime, $c_{T}/c_{B}$ 
decreases monotonically with increasing deviation from particle-hole symmetry. Since $c_{T}/c_{B}$ is independent of the definition of Kondo scale used, 
it could be a useful quantity to quantify the degree of correlations in a quantum dot ($U/\Delta$) and the degree of departure
from particle-hole symmetry for specific gate voltages.

\subsubsection{Comparison with SPT}

\begin{figure}
\includegraphics[width=\linewidth,clip]{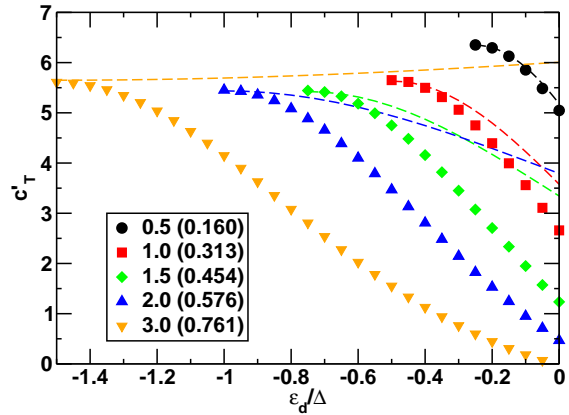}

\caption
{
  {\em (Color online)}
  $c'_{T}$ vs $\varepsilon_{d}/\Delta$ for several $U/\Delta$ calculated within NRG (symbols) and SPT (lines). 
  Legend: column one $U/\Delta$, column two $\tilde{u}$. 
  $c'_{T}$ is defined in  Eq.~(\ref{ct-expression}) using the Kondo scale in Eq.~(\ref{gt-spt}), and the corresponding 
  NRG result uses the same Kondo scale for the purposes of this comparison.
  NRG parameters were for $\Lambda = 4$ with an  energy cut-off $e_{c}(\Lambda=4)=30$ and  $n_{z}=2$.
  \label{fig-ct-compare}
}
\end{figure}

\begin{figure}
\includegraphics[width=\linewidth,clip]{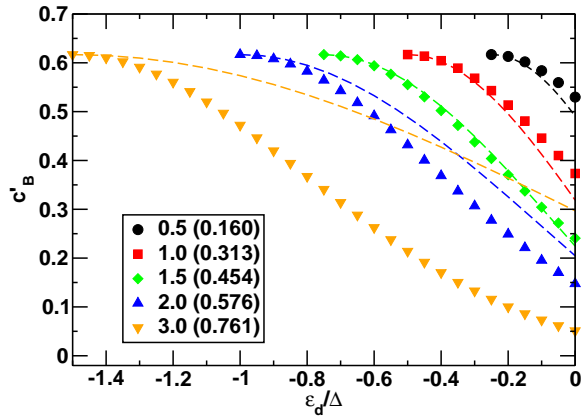}
\caption
{
  {\em (Color online)}
  $c_{B}$ vs $\varepsilon_{d}/\Delta$ for several $U/\Delta$ calculated within NRG (symbols) and SPT (lines). 
  Legend: column one $U/\Delta$, column two $\tilde{u}$. 
  $c'_{B}$ is defined in  Eq.~(\ref{cb-expression}) using the Kondo scale in Eq.~(\ref{gb-spt}), and the corresponding 
  NRG result uses the same Kondo scale for the purposes of this comparison.
  NRG parameters were for $\Lambda = 4$ with an  energy cut-off $e_{c}(\Lambda=4)=30$ and  $n_{z}=2$.
  \label{fig-cb-compare}
}
\end{figure}

Figure~\ref{fig-ct-compare} compares SPT results for the local level dependence of $c_{T}'$, as defined in Eqs.~(\ref{ct-expression}) and (\ref{gt-spt}),
with correspondingly defined quantities in NRG. Figure~\ref{fig-cb-compare} shows a similar comparison for the quantity
$c_{B}'$ defined in Eqs.~(\ref{cb-expression}) and (\ref{gb-spt}). 
We see in both  cases, that agreement between NRG and SPT, holds for 
$\tilde{\varepsilon}_{d}\equiv (\varepsilon_{d}+U/2)/\Delta \lesssim 0.25$, which is consistent with the SPT calculations carried out
to order $\tilde{u}^{2}\tilde{\varepsilon}^{2}_{d}$. For larger deviations from the symmetric point and with increasing Coulomb interactions,
we see an increasing deviation of the SPT results from the NRG calculations. In contrast to the NRG calculation, we also see that the
SPT result for $c_{T}'$ ceases to decrease monotonically with $\tilde{\varepsilon}_{d}$ for $U/\Delta\gtrsim 3$ (corresponding to a renormalized
Coulomb interaction $\tilde{u}\gtrsim 0.76$). On the other hand, the SPT result for $c_{B}'$ decreases monotonically with increasing $\tilde{\varepsilon}_{d}$
as in the corresponding NRG result. Although we show comparisons also in the region $\tilde{\varepsilon}_{d}\gg 1$, by
construction the SPT calculation is perturbative in $\tilde{\varepsilon}_{d}$ and agreement can only be expected in the limit $\tilde{\varepsilon}_{d}\ll 1$, which we
find. We also expect that the range of agreement between NRG and SPT in both $\tilde{u}$ and $\tilde{\varepsilon}_{d}$ can be increased by going to
higher-order (see discussion at the end of Sec.~\ref{subsec:retarded self-energy}), however, this lies beyond the scope of this paper. 

\section{Conclusions}
\label{sec:conclusions}
In this paper we investigated deviations from the universal Kondo scaling in the linear conductance of a correlated
quantum dot due to a finite level asymmetry (i.e., deviation of gate voltage from mid-valley) and a finite local Coulomb
repulsion (i.e., finite charging energy). In particular, we determined the behavior of the
coefficients $c_{T}$ and $c_{B}$  as a function of $\varepsilon_{d}$ and $U$ within NRG and compared these with 
results from SPT,\cite{Munoz2013} finding good agreement for all $U$ at the symmetric point and reasonable agreement for 
$\tilde{\varepsilon}_{d}=(\varepsilon_{d}+U/2)/\Delta\lesssim 0.25$ away from the symmetric point. Both $c_{T}$ and $c_{B}$ are 
monotonically decreasing functions of the deviation $\tilde{\varepsilon}_{d}$ from the symmetric point $\tilde{\varepsilon}_{d}=0$
for all $U$ and an exact Fermi liquid expression for $c_{B}$ has been given which is valid for any $U$ and $\varepsilon_{d}$.
In particular, the coefficients $c_{T}$ and $c_{B}$ become negative on entering the mixed valence regime, signaling the onset of thermally 
activated transport which becomes pronounced in the empty orbital limit $n_{d}\approx 0$. 

For the mid-valley conductance, we also determined the ratio of the conductance to susceptibility Kondo scales $T_{\rm K}^{\rm expt}/T_{0}^{(s)}$, 
allowing us to relate our results for $c_{T}$ and $c_{B}$ in terms of $T_{0}^{(s)}$, to the measured coefficients $c_{T}^{\rm expt}$ and 
$c_{B}^{\rm expt}$ in terms of $T_{\rm K}^{\rm expt}$. While for quantum dots with $U/\Delta\gtrsim 6$, the difference
between the two sets of coefficients is a constant factor of order unity 
(e.g., $c_{T,B}^{\rm expt}/c_{T,B}=(T_{\rm K}^{\rm expt}/T_{0}^{(s)})^2 \approx 1.08$ for $U/\Delta \gg  6$), for quantum dots with 
$U/\Delta\lesssim 6$ this difference becomes significant and should be carefully taken into account in detailed comparisons
of theory with experiment. We expect this to be particularly important for 
semiconducting quantum dots since $U/\Delta$ is tunable to smaller values in these systems. 

\begin{acknowledgments}
T.A.C and L.M. thank A. Weichselbaum for some useful comments on this work and acknowledge supercomputer support by the John von
Neumann institute for Computing (J\"ulich).E.M. and S.K. acknowledge support by the Comisi\'{o}n Nacional de 
Investigaci\'{o}n Cient\'{i}fica y Tecnol\'{o}gica (CONICYT), grant No. 11100064 and the German Academic Exchange Service (DAAD) under grant No. 52636698.
\end{acknowledgments}

\appendix
\section{Alternative derivation of the Campo discretization}
\label{appendix:campo}
In this appendix we give a derivation of the discretization scheme of Ref.~\onlinecite{Campo2005}, 
following the procedure for general energy dependent hybridization functions of Bulla {\it et al.} in 
Refs.~\onlinecite{Bulla1997} and \onlinecite{Bulla2008}, 
which has been used for the NRG calculations of the conductance in this paper.

We start with the single-channel Anderson impurity model, given by
\begin{align*}
 H = & H_{imp} +
          \sum_{k,\sigma} \epsilon_{k,\sigma} c^\dagger_{k,\sigma} c_{k,\sigma} \nonumber \\
          &+  \sum_{k,\sigma} V_k (f^\dagger_\sigma c_{k,\sigma} + c^\dagger_{k,\sigma} f_\sigma), 
\end{align*}
which may be written in the energy representation as \cite{Bulla1997}
\begin{align}
\label{def:cont_ham}
 H = & H_{imp} + \sum_\sigma \int_{-D_-}^{D_+} h(\epsilon) (f^\dagger_\sigma a_{\epsilon,\sigma} + a^\dagger_{\epsilon,\sigma} f_\sigma) d \epsilon \nonumber \\
& + \sum_\sigma \int_{-D_-}^{D_+} g(\epsilon) a^\dagger_{\epsilon,\sigma} a_{\epsilon,\sigma} d \epsilon.
\end{align}
Here, $a_{\epsilon,\sigma}$ and $a_{\epsilon',\sigma'}^{\dagger}$ obey the standard anticommutation relations 
$\{ a_{\epsilon,\sigma},a_{\epsilon',\sigma'}^{\dagger} \}=\delta_{\sigma,\sigma'}\delta(\epsilon-\epsilon')$,  $g(\epsilon)$ is the  
dispersion, $h(\epsilon)$ is the hybridization amplitude, and $\pm D_{\pm}$ 
are the upper/lower conduction electron band edges. The model (\ref{def:cont_ham}) is characterized by the hybridization function 
$\Delta(\omega)=\sum_k |V_k|^2 \delta(\omega - \epsilon_k)$. As shown in Ref.~\onlinecite{Bulla1997}, its energy dependence may be 
distributed arbitrarily over the functions $g(\epsilon)$ and $h(\epsilon)$, as long as the following condition is satisfied 
\begin{equation}
\label{def:delta}
  \Delta(\omega) = \pi \frac{d \epsilon(\omega)}{d \omega} h(\epsilon(\omega))^2,
\end{equation}
where $\epsilon(\omega)$ is the inverse function of the dispersion $g(\epsilon)$, i.e.,
\begin{equation*}
  g(\epsilon(\omega)) = \omega.
\end{equation*}

Our starting point is the observation by Campo {\it et al.} \cite{Campo2005} that a linear discretization of the conduction band with 
a Fourier basis in the discrete intervals leads to a correct estimate for $\Delta(\omega)$, whereas a Fourier 
decomposition on a logarithmic scale as suggested by Krishna-Murthy et al. in Ref.~\onlinecite{KWW1980a} 
systematically underestimates $\Delta(\omega)$ (or equivalently the conduction electron density of states $\rho(\omega)$ since
$\Delta(\omega)=\pi\rho(\omega)V^{2}$ for a constant hybridization matrix element $V_{k}=V$). 
This underestimation results in an effective hybridization function $\tilde{\Delta}(\omega)=\Delta(\omega)/A_{\Lambda}$ 
where the factor $A_{\Lambda}= \frac{\ln \Lambda}{2}\frac{1+\Lambda^{-1}}{1-\Lambda^{-1}}>1$
is due the discretization and $\Lambda>1$ is the band discretization parameter.
\footnote{For example, $A_{\Lambda}=1.41,1.15,1.04,1.01$ for $\Lambda=10,4,2,1.5$.} 
While this effect may be corrected ``manually'' for each $\Lambda$, it is clearly advantageous to have a 
built-in procedure within the NRG that does this automatically. Campo et al. accomplished this within 
a logarithmic discretization scheme by using a Fourier decomposition in terms of non-orthogonal basis functions. 
As in the case of the linear grid, this correctly estimated $\Delta(\omega)$. Motivated by this, we provide here 
an alternative derivation of this discretization scheme following the procedure of Bulla {\it et al.} in 
Refs.~\onlinecite{Bulla1997} and \onlinecite{Bulla2008} for general $\Delta(\omega)$.

We consider the following set of orthonormal Fourier functions in each interval of a linear grid,
\begin{equation*}
\psi_{n,p}(\eta) = 
\begin{cases}
 e^{- 2 \pi i p \eta}, &\text{if } \eta \in [n,n+1], n=-1,0,1,\dots\\
  0, & \text{otherwise.}
\end{cases}
\end{equation*}
The inverse functions are given by $\Psi_{n,p}(\eta) = \psi_{n,p}^{*}(\eta)$ fulfilling the usual orthonormality condition:
\begin{equation}
\label{def:fourier}
 \int_{-\infty}^{\infty} \psi_{n,p}(\eta) \Psi_{n',p'}(\eta) d \eta = \delta_{n, n'} \delta_{p, p'}
\end{equation}
We will transform this relation to a logarithmic grid such that $[n,n+1]$ will be transformed to $ D_{+}[\Lambda^{-n-z-1},\Lambda^{-n-z}]$ for $n=0,1,\dots$. 
The first interval $[-1,0]$ is special and transforms to the first logarithmic interval containing the band edge, i.e. to $D_{+}[\Lambda^{-z},1]$. 
One possible choice, the obvious one, 
is $\epsilon = D_+ \Lambda^{-\eta-z}\leftrightarrow \eta(\epsilon)=-\ln |\epsilon/D_{+}|/\ln\Lambda -z, n=0,1,\dots$ 
($\epsilon=D_{+}\Lambda^{-z(\eta+1)}, n=-1$) , but other choices are possible for defining 
the transformation between linear and logarithmic grids (and hence $\eta(\epsilon)$).
\footnote{A different choice will be made later in Eq.~(\ref{def:eta}). At present, the only requirement on $\eta(\epsilon)$ is
that $\eta(\pm D_{\pm}\Lambda^{-n-z})=n$ and $\eta(\pm D_{\pm}\Lambda^{-n-1-z})=n+1$, i.e., the boundaries of $[n,n+1]$ map onto the
mesh points of the logarithmic grid.} 
Thus, for $n=0,1,\dots$ we 
have\footnote{For $n=-1$ a similar expression applies on using $\epsilon=D_{+}\Lambda^{-z(\eta+1)}$.}
\begin{align}
&\int_{0}^{\infty} \underbrace{\frac{1}{\left| \epsilon \right| \ln \Lambda} \psi_{n,p}(-\frac{\ln \frac{\left| \epsilon \right|}{D_+}}{\ln \Lambda} -z)}_{= \phi^+_{n,p}(\epsilon)/c_n^+} \underbrace{\Psi_{n',p'}(-\frac{\ln \frac{\left| \epsilon \right|}{D_+}}{\ln \Lambda} -z)}_{= c_n^+ \cdot \Phi_{n',p'}^+(\epsilon) } d \epsilon\nonumber\\ 
&= \delta_{n, n'} \delta_{p, p'}\label{def:phi_plus} 
\end{align}
For the expansion of the negative part of the band we use $\epsilon = -D_- \Lambda^{-\eta-z}$ and do not reverse the integration boundaries yielding the same function inside the integration. 
\begin{align}
 &\int_{-\infty}^{0} \underbrace{\frac{1}{\left| \epsilon \right| \ln \Lambda} \psi_{n,p}(-\frac{\ln \frac{\left| \epsilon \right|}{D_-}}{\ln \Lambda} -z)}_{= \phi^-_{n,p}(\epsilon)/c_n^-} \underbrace{\Psi_{n',p'}(-\frac{\ln \frac{\left| \epsilon \right|}{D_-}}{\ln \Lambda} -z)}_{= c_n^- \cdot \Phi_{n',p'}^-(\epsilon) } d \epsilon\nonumber \\ 
&= \delta_{n, n'} \delta_{p, p'}\label{def:phi_minus}
\end{align}
The normalization factor $c_n^\pm$ can be distributed freely between the new basis $\phi^\pm_{n,p}$ and its inverse $\Phi^\pm_{n,p}$.
$a_{\epsilon,\sigma}$ is expressed in terms of the new basis:
\begin{equation*}
 a_{\epsilon, \sigma} = \sum_{n,p} a_{n,p,\sigma} \phi_{n,p}^+(\epsilon) + b_{n,p,\sigma} \phi_{n,p}^-(\epsilon)
\end{equation*}
where
\begin{align*}
a_{n,p,\sigma} & = \int^{+n} a_{\epsilon, \sigma} \Phi_{n,p}^+(\epsilon) d \epsilon \\
b_{n,p,\sigma} & = \int^{-n} b_{\epsilon, \sigma} \Phi_{n,p}^-(\epsilon) d \epsilon
\end{align*}
where we defined
\begin{equation*}
  \int^{+n} = \int_{D_+ \Lambda^{-n-z-1}}^{D_+ \Lambda^{-n-z}}, \quad \int^{-n} = \int_{- D_- \Lambda^{-n-z}}^{- D_- \Lambda^{-n-z-1}}
\end{equation*}
Evaluating the anticommutator $\left\{ a_{n,p,\sigma},a_{n',p',\sigma'}^\dagger \right\}$ we find
\begin{align*}
 &&&\left\{ a_{n,p,\sigma},a_{n',p',\sigma'}^\dagger \right\} \\
& = &&\int^{+n} a_{\epsilon, \sigma} \Phi_{n,p}^+(\epsilon) d \epsilon \int^{+n'} a_{\epsilon', \sigma'}^\dagger {\Phi_{n',p'}^+}^*(\epsilon') d \epsilon' \\
 & &&+ \int^{+n'} a_{\epsilon', \sigma'}^\dagger {\Phi_{n',p'}^+}^*(\epsilon') d \epsilon \int^{+n} a_{\epsilon, \sigma} \Phi_{n,p}^+(\epsilon) d \epsilon \\
 & =&& \delta_{n,n'} \int^{+n} \int^{+n} [a_{\epsilon, \sigma}, a_{\epsilon', \sigma'}^\dagger] \Phi_{n,p}^+(\epsilon) {\Phi_{n,p'}^+}^*(\epsilon') d \epsilon d \epsilon' \\
 & = &&\delta_{n,n'} \delta_{\sigma,\sigma'} \int^{+n} \Phi_{n,p}^+(\epsilon) {\Phi_{n,p'}^+}^*(\epsilon) d \epsilon \\
 & =&& \delta_{n,n'} \delta_{\sigma,\sigma'} \int^{+n} \frac{1}{\left| c_n^{+} \right|^2} \mathrm{e}^{2 \pi i (p-p')(\frac{\ln \frac{\left| \epsilon \right|}{D_\pm}}{\ln \Lambda} + z)}
\end{align*}
with an analogous expression for $\left\{ b_{n,p,\sigma},b_{n',p',\sigma'}^\dagger \right\}$. Setting 
$\left\{ a_{n,p,\sigma},a_{n,p,\sigma}^\dagger \right\}=\left\{ b_{n,p,\sigma},b_{n,p,\sigma}^\dagger \right\}=1$ fixes the constants
$c_n^{\pm}$  in Eqs.~(\ref{def:phi_plus}) and (\ref{def:phi_minus}), leading to
\begin{align*}
 \left|c_n^\pm\right|^2 &= D_\pm \Lambda^{-n-z}(1-\Lambda^{-1}) \quad = d_n^\pm \\
  &(= D_\pm(1 - \Lambda^{-z}) \quad \text{for} \, n = -1)
\end{align*}
and
\begin{equation*}
\left\{ a_{n,p,\sigma},a_{n',p',\sigma'}^\dagger \right\}= \delta_{n,n'} \delta_{\sigma,\sigma'}
\begin{cases}
 1, &\text{if } p = p'\\
  \frac{\ln \Lambda}{2 \pi i (p - p') + \ln \Lambda}, & \text{otherwise,}
\end{cases}
\end{equation*}
with an analogous expression for $\left\{ b_{n,p,\sigma},b_{n',p',\sigma'}^\dagger \right\}$.
Thus, only for the continuum limit $\Lambda \rightarrow 1$ is the above an orthonormal basis for all $p,p'$.\cite{Campo2005} 
However, as we show below, an approximate discretized Hamiltonian can be formulated in terms of the orthonormal subset of $p=0$ states only, 
within which the NRG calculation is carried out.  We show that for general $\Delta(\omega)$, (i), only $p=0$ states couple to the impurity, 
and, (ii), off-diagonal terms in $p,p'$ can always be eliminated from the Hamiltonian by a suitable choice of the function $\eta(\epsilon)$ 
relating the linear to the logarithmic discretization $\epsilon=\pm D_{\pm} \Lambda^{-\eta(\epsilon)-z}$. 

With the new basis functions we follow the derivation of Bulla et al. in Ref.~\onlinecite{Bulla2008},
reformulating first the hybridization part of Eq.~(\ref{def:cont_ham})
\begin{align*}
 \int_{-D_-}^{D_+} h(\epsilon) a_{\epsilon,\sigma} d \epsilon & = \sum_{n,p} a_{n,p,\sigma} \int^{+n} h(\epsilon) \phi_{n,p}^+(\epsilon) d \epsilon \\
&+ \sum_{n,p} b_{n,p,\sigma} \int^{-n} h(\epsilon) \phi_{n,p}^-(\epsilon) d \epsilon
\end{align*}

The requirement that the hybridization only couples to the $p=0$ terms can be satisfied by choosing 
$h(\epsilon) \propto \Phi_{n,0}^{\pm}(\epsilon) = \frac{1}{\sqrt{d_n}}$, which by Eqs.~(\ref{def:phi_plus}) and (\ref{def:phi_minus}) implies that $p\neq 0$ do not hybridize. 
Therefore we can choose the same $h(\epsilon)$ as in Ref.~\onlinecite{Bulla2008} (i.e. a step function in the discrete intervals),
\begin{equation}
 h(\epsilon)^2 \equiv {h_{n}^{\pm}}^{2} = \frac{1}{d_{n}^{\pm}}\int^{\pm n} \frac{1}{\pi}\Delta(\omega)d  \omega,\label{eq:h-disc}
\end{equation}
for $D_\pm \Lambda^{-n-z}<\pm \epsilon <D_\pm \Lambda^{-n-1 -z}$. This choice guarantees that 
$\epsilon(\pm D_\pm \Lambda^{-n-z})=\pm D_\pm \Lambda^{-n-z}$ 
[proved by using Eqs.~(\ref{def:delta}) and (\ref{eq:h-disc})] and that the dispersion is linear 
at the grid points $g(\pm D_\pm \Lambda^{-n-z})=\pm D_\pm \Lambda^{-n-z}$. The first part of the hybridization may be written as
\begin{align*}
  \sum_\sigma \int_{-D_{-}}^{+D_{+}} h(\epsilon) f^\dagger_\sigma a_{\epsilon,\sigma} d \epsilon &= \frac{1}{\sqrt{\pi}}  
\sum_{n} f^\dagger_\sigma  \left(\gamma_n^+ a_{n,0,\sigma} + \gamma_n^- b_{n,0,\sigma}\right) \\
&\equiv \sqrt{\frac{\xi_{0}}{\pi}}\sum_{\sigma}f_{\sigma}^{\dagger}f_{0\sigma}
\end{align*}
where $ {\gamma_n^\pm}^2 = \int^{\pm n} \Delta(\omega) d \omega,$
and the conduction electron Wannier orbital at the impurity site is defined as
$
f_{0\sigma}=\frac{1}{\sqrt{\xi_{0}}}\sum_{n} \gamma_n^+ a_{n,0,\sigma} + \gamma_n^- b_{n,0,\sigma}, \\
$
with 
$
\xi_{0} =\sum_{n} \left({\gamma_n^+}\right)^{2} + \left({\gamma_n^-}\right)^{2}. \\
$

Next, we reformulate the conduction electron kinetic energy term:
\begin{align*}
 \int_{-D_-}^{D_+} g(\epsilon) a_{\epsilon, \sigma}^\dagger a_{\epsilon, \sigma} d\epsilon &\\
 = \sum_{n} \sum_{p,p'}  &a_{n,p,\sigma}^\dagger a_{n,p',\sigma} \int^{+n} g(\epsilon) \phi_{n,p}^+(\epsilon) {\phi_{n,p'}^+}^*(\epsilon) d \epsilon \\
 +&b_{n,p,\sigma}^\dagger b_{n,p',\sigma} \int^{-n} g(\epsilon) \phi_{n,p}^-(\epsilon) {\phi_{n,p'}^-}^*(\epsilon) d \epsilon \nonumber \\
 = \sum_{n} \sum_{p,p'} &a_{n,p,\sigma}^\dagger a_{n,p',\sigma} \xi_{n,p,p'}^+ +b_{n,p,\sigma}^\dagger b_{n,p',\sigma} \xi_{n,p,p'}^- 
\end{align*}
\begin{align*}
 \xi_{n,p,p'}^\pm &= \int^{\pm n} g(\epsilon) \phi_{n,p}^\pm(\epsilon) {\phi_{n,p'}^\pm}^*(\epsilon) d \epsilon \\
 & = \int^{\pm n} \omega \Delta(\omega) \frac{1}{\pi h(\epsilon)^2}\phi_{n,p}^\pm(\epsilon) {\phi_{n,p'}^\pm}^*(\epsilon) d \omega \\
 & = \frac{\int^{\pm n} \omega \Delta(\omega)d_n^{\pm}\phi_{n,p}^\pm(\epsilon(\omega)) {\phi_{n,p'}^\pm}^*(\epsilon(\omega)) d \omega}{\int^{\pm n} \Delta(\omega) d\omega}\\
& = \frac{\int^{\pm n} \frac{\omega \Delta(\omega){d_n^{\pm}}^2}{\left(\left|\epsilon(\omega)\right| \ln \Lambda\right)^2} \mathrm{e}^{2 \pi i (p -p')(\frac{\ln \frac{\left| \epsilon(\omega) \right|}{D_\pm} }{\ln \Lambda}+z)}d \omega}{ \int^{\pm n} \Delta(\omega) d\omega}
\end{align*}
For $\Delta(\omega) = \Delta_0$ we obtain $\epsilon(\omega) = \omega$ and only $\xi_{n,p=p'}^\pm = \pm \frac{d_{n}^{\pm}}{\ln \Lambda}$ 
are unequal to zero and agree with the result of Campo and Oliveira. Thus, the impurity, which by construction couples only to the 
$p=0$ state via the hybridization term, is completely decoupled from the $p\neq 0$ states. We now show that the same can be achieved 
for a general $\Delta(\omega)$ by a suitable choice of $\eta(\epsilon)$.

Following the same derivation as above, but substituting 
\begin{align}
\eta(\epsilon) = \int_{k_n^{(2)}}^{\epsilon} \frac{k_n^{(1)}}{g(\epsilon')}d \epsilon'\label{def:eta}
\end{align}
in Eq.~(\ref{def:fourier}) leads to diagonal $\xi_{n,p,p'}$ for an arbitrary $\Delta(\omega)$.
$k_n^{(1)}$ and $k_n^{(2)}$ are given by the boundary conditions $\int_{k_n^{(2)}}^{\pm D_\pm \Lambda^{-n-z}} \frac{k_n^{(1)}}{g(\epsilon')} d\epsilon̈́' = n$ and
$\int_{k_n^{(2)}}^{\pm D_\pm \Lambda^{-n-z-1}} \frac{k_n^{(1)}}{g(\epsilon')} d\epsilon̈́' = n+1$.
$c_n^\pm$ and $\gamma^\pm_n$ remain unchanged but
\begin{align*}
   \xi_{n,p,p'}^\pm &= \int^{\pm n} g(\epsilon) \phi_{n,p}^\pm(\epsilon) {\phi_{n,p'}^\pm}^*(\epsilon) d \epsilon \\
   & = \int^{\pm n} \frac{ d_{n}^{\pm} {k_n^{(1)}}^2}{g(\epsilon)} \mathrm{e}^{2 \pi i (p - p') \int_{k_n^{(2)}}^{\epsilon} \frac{k_n^{(1)}}{g(\epsilon')}d \epsilon'} d \epsilon\\
  & = \mp k_n^{(1)} d_{n}^{\pm} \delta_{p,p'}.
\end{align*}
$k_n^{(1)}$ can be obtained by taking the difference of the boundary conditions (as defined above), and using 
$\epsilon(\pm D_\pm \Lambda^{-n-z})=\pm D_\pm \Lambda^{-n-z}$ and Eq.~(\ref{def:delta}), 
$$k_n^{(1)} = \frac{\int^{\pm n} \Delta(\omega) d\omega}{\mp d_{n}^{\pm} \int^{\pm n} \frac{\Delta(\omega)}{\omega} d\omega}$$
As in Ref.~\onlinecite{Campo2005} the resulting $\xi_{n,p=p'}^\pm\equiv \xi_{n,p=p'}^{\pm}(C)$ are given by
\footnote{Note that the derivation in Ref.~\onlinecite{Campo2005} was actually for the two-impurity Anderson model, where the energy dependent
 couplings arise via the non-trivial $k$-dependence of the hybridization term $H_{\rm hyb}=V\sum_{\vec{k},j=1,2}(e^{i\vec{k}\dot\vec{R}_{j}}c^{\dagger}_{\vec{k}}d_{j}+H.c.)$.}

 \begin{equation}
  \xi_{n,p=p'}^\pm(C)  = \frac{\int^{\pm n} \Delta(\omega) d\omega}{ \int^{\pm n} \frac{\Delta(\omega)}{\omega} d\omega}\label{eq:campo-xi}
 \end{equation}
The corresponding result, denoted by $\xi_{n,p=p'}^\pm=\xi_{n,p=p'}^{\pm}(B)$, in the usual logarithmic discretization scheme is given by\cite{Bulla1997,Sakai1994}
 \begin{equation}
  \xi_{n,p=p'}^\pm(B)  = \frac{\int^{\pm n} \omega\Delta(\omega) d\omega}{ \int^{\pm n} \Delta(\omega) d\omega}\label{eq:bulla-xi}.
 \end{equation}
Evaluating (\ref{eq:campo-xi}-\ref{eq:bulla-xi}) for a flat band with $D_{+}=D_{-}=1$, gives 
\begin{align*}
\xi_{n,p=p'}^\pm(C) &= \pm\frac{1}{2}(1+\Lambda^{-1})\Lambda^{-n-z}/A_{\Lambda},\,\, n=0,1,\dots\\
\xi_{n,p=p'}^\pm(C) &= \pm\frac{1}{2}(1+{\left(\Lambda^z\right)}^{-1})/A_{\Lambda^{z}},\,\, n=-1\\ 
\xi_{n,p=p'}^\pm(B) &= \pm\frac{1}{2}(1+\Lambda^{-1})\Lambda^{-n-z},\,\, n=0,1,\dots\\ 
\xi_{n,p=p'}^\pm(B) &= \pm\frac{1}{2}(1+{\left(\Lambda^z\right)}^{-1}),\,\, n=-1
\end{align*}
We see that $\xi_{n,p=p'}^\pm(B)/\xi_{n,p=p'}^\pm(C)$ is given by the factor $A_{\Lambda}$ ($A_{\Lambda^z}$ for $n=-1$), 
indicating that the Campo discretization achieves the correct estimate for $\Delta(\omega)$ via a reduction of the effective
bandwidth of the discretized model. For energies close to the band edge, where the Campo discretization gives a different
correction to the desired one  ($A_{\Lambda^z}$ instead of $A_{\Lambda}$) further corrections are needed.\cite{Zitko2009b}

\section{SPT calculation of $c_{B}$}
\label{appendix:cb calculation}
We describe the system by the single impurity Anderson Hamiltonian,
$\hat{H} = \hat{H}_c + \hat{H}_d + \hat{H}_{d-c}$,
where we defined
\begin{eqnarray}
\hat{H}_c &&= \sum_{\lambda = L, R}
\sum_{k,\sigma}\epsilon_{k\lambda}
\hat{c}_{k\lambda\sigma}^{\dagger}\hat{c}_{k\lambda\sigma}\label{eqB1}\\
\hat{H}_d &&= \sum_{\sigma} E_{d,\sigma}
\hat{d}_{\sigma}^{\dagger}\hat{d}_{\sigma} + U\left(\hat{d}_{\uparrow}^{\dagger}\hat{d}_{\uparrow} - \frac{1}{2}\right)\left(\hat{d}_{\downarrow}^{\dagger}\hat{d}_{\downarrow} - \frac{1}{2}\right) - \frac{U}{4}\nonumber\\
\hat{H}_{d-c} &&= \sum_{\lambda = L,
R}\sum_{k,\sigma}\left(V_{k\lambda}\hat{d}_{\sigma}^{\dagger}\hat{c}_{k\lambda\sigma}
+V_{k\lambda}^{*}\hat{c}_{k\lambda\sigma}^{\dagger}\hat{d}_{\sigma}\right).\nonumber
\end{eqnarray}

Here, $\hat{H}_c$ is the single-band Hamiltonian for conduction electrons at the metallic leads. $H_{d}$ is the Hamiltonian
 for localized quasi-particle states at the dot, which includes Coulomb interaction.  $\hat{H}_{d-c}$ represents the coupling between the dot and the leads.
We have defined the spin-dependent local energy level $E_{d\sigma} = E_d - \sigma b$, with $E_{d} = \epsilon_{d} + U/2$ a small parameter to capture deviations
from the particle-hole (p-h) symmetric condition $\epsilon_{d} = -U/2$, and $b = g\mu_{B}B/2$. 
We build up an SPT
calculation starting from a reference system which 
is interacting ($U\ne 0$), particle-hole symmetric ($E_{d} = \epsilon_{d} + U/2 =0$), and
in the absence of an external magnetic field ($B=0$). 

\subsection{The reference system}
The reference system ($E_{d}=0$, $B=0$) self-energy was derived in detail in \cite{Munoz2013},
and is given by the matrix form
\begin{eqnarray*}
{\mathbf{\Sigma}}_{\sigma,\omega} = \left[\begin{array}{cc}\Sigma^{--}_{\sigma,\omega} & -\Sigma^{-+}_{\sigma,\omega}\\
-\Sigma^{+-}_{\sigma,\omega} & \Sigma^{++}_{\sigma,\omega}\end{array} \right].
\label{eqB2}
\end{eqnarray*}
The local Green's function for the reference system is given by the matrix
\begin{eqnarray*}
{\mathbf{g}}_{\sigma,\omega} = \left[\begin{array}{cc}g^{--}_{\sigma,\omega} & g^{-+}_{\sigma,\omega}\\
g^{+-}_{\sigma,\omega} & g^{++}_{\sigma,\omega}\end{array} \right],
\end{eqnarray*}
with components satisfying \cite{Munoz2013}
\begin{eqnarray}
g^{--}_{\sigma,\omega} &&= [1 - F(\omega,T,V)]g^{r}_{\sigma,\omega} +
F(\omega,T,V)g^{a}_{\sigma,\omega}\nonumber\\
g^{-+}_{\sigma,\omega} &&= -F(\omega,T,V)[g^{r}_{\sigma,\omega}
 - g^{a}_{\sigma,\omega}]\label{eqB4}
\\
g^{+-}_{\sigma,\omega} &&= [1-F(\omega,T,V)][g^{r}_{\sigma,\omega}
 - g^{a}_{\sigma,\omega}]\nonumber\\
g^{++}_{\sigma,\omega} &&= -[1-F(\omega,T,V)]g^{a}_{\sigma,\omega} - F(\omega,T,V)g^{r}_{\sigma,\omega}.\nonumber
\end{eqnarray}
Here, the effective local nonequilibrium distribution function is shown \cite{Munoz2013} to be
\begin{equation*}
F(\omega,T,V) = \frac{\Delta_{L}f_{L} + \Delta_{R}f_{R}-(i/2)\Sigma_{\omega}^{-+}}{\Delta_{L} + \Delta_{R} - {\rm{Im}}\Sigma_{\omega}^{r}}.
\end{equation*}
The retarded component of the Green's function is given by
\begin{equation*}
g_{\sigma\omega}^{r} = \left(\omega + i\Delta - \Sigma^{r}_{\omega} \right)^{-1},
\end{equation*}
with $g_{\sigma\omega}^{a}=\left[g_{\sigma\omega}^{r}\right]^{*}$, and $\Delta = \Delta_{L}+\Delta_{R}$.
It is shown in Ref.\cite{Munoz2013} that the self-energy components for the reference system
satisfy a similar set of relations as Eq.(\ref{eqB4}), in particular with
\begin{eqnarray*}
\Sigma^{-+}_{\omega} &=& -2i{\rm{Im}}\Sigma_{\omega}^{r}F(\omega,T,V)\\
\Sigma^{+-}_{\omega} &=& 2i{\rm{Im}}\Sigma_{\omega}^{r}[1-F(\omega,T,V)].
\end{eqnarray*}
The (spin-independent) retarded self-energy of the reference system was calculated in detail
in \cite{Munoz2013}, and is given by
\begin{eqnarray}
\Sigma^{r}_{\omega} &=& (1 - \tilde{\chi}_{++})\omega - i\Delta\frac{\tilde{\chi}_{+-}^{2}}{2}\left[\left(\frac{\omega}{\Delta}\right)^{2} + \left(\frac{\pi T}{\Delta}\right)^{2}\right.\nonumber\\
&&\left.+\zeta\left(\frac{e V}{\Delta} \right)^{2}- \tilde{\chi}_{++}^{2}\frac{\zeta}{3}\left(\frac{\pi T eV}{\Delta^{2}}\right)^{2}\right].
\label{eqB5}
\end{eqnarray}
with $\Sigma^{a}_{\omega} = \left(\Sigma_{\omega}^{r}\right)^{*}$. In Eq.(\ref{eqB5}), the "odd" component
of the spin susceptibility is directly related to the four-point vertex $\Gamma_{0}(U)$ defined in Eq.(\ref{eqvertex}) by the relation $\tilde{\chi}_{+-}(U) = \Gamma_{0}/(\pi\Delta)$. 

\subsection{SPT formulation}
A coherent-states path-integral representation of the model of Eq.(\ref{eqB1}) in terms of
 Grassmann fields can be obtained following the standard construction.
 On the Keldysh contour, we introduce
 $\hat{\psi}_{k\lambda\sigma}(t)^{\dagger} = \left(c_{k\lambda\sigma}^{-}(t),c_{k\lambda\sigma}^{+}(t)\right)^{\dagger}$ 
 and $\hat{\Phi}(t)^{\dagger} = \left(d_{\sigma}^{-}(t),d_{\sigma}^{+}(t)\right)^{\dagger}$, where the indexes $\pm$ refer to the
time-ordered (-) and anti-time-ordered (+) paths, while $\lambda = \{L, R\}$ labels the two different leads.

The resulting non-equilibrium generating functional for the model of  Eq.(\ref{eqB1}) is \cite{Munoz2013,Kamenev2004},
\begin{eqnarray}
Z = \int\mathcal{D}[\hat{\psi}^{\dagger},\hat{\psi}]\mathcal{D}[\hat{\Phi}^{\dagger},\hat{\Phi}]
e^{iS[\hat{\psi}^{\dagger},\hat{\psi},\hat{\Phi}^{\dagger},\hat{\Phi}]}.
\label{eqB9}
\end{eqnarray}

Since the action in Eq.(\ref{eqB9}) is Gaussian in the $\hat{\psi}_{k\lambda\sigma}^{\dagger}(t)$, $\hat{\psi}_{k\lambda\sigma}(t)$
Grassmann fields, we integrate those in the partition function Eq.(\ref{eqB9}) to obtain, in the frequency-space representation,
\begin{eqnarray}
Z = \int\mathcal{D}[\hat{\Phi}^{\dagger}_{\sigma\omega},\hat{\Phi}_{\sigma\omega}]e^{iS[\hat{\Phi}^{\dagger}_{\sigma\omega},\hat{\Phi}_{\sigma\omega}]}.
\label{eqB10}
\end{eqnarray}
In Eq.(\ref{eqB10}), we have defined the effective action as
\begin{eqnarray*}
i S[\hat{\Phi}^{\dagger}_{\sigma\omega},\hat{\Phi}_{\sigma\omega}] &=& i S_{U}[\hat{\Phi}^{\dagger}_{\sigma\omega},\hat{\Phi}_{\sigma\omega}]\\
&&-
i\int_{-\infty}^{+\infty}\frac{d\omega}{2\pi}\sum_{\sigma}\hat{\Phi}^{\dagger}_{\sigma\omega}E_{d\sigma}\hat{\sigma}_{3}\hat{\Phi}_{\sigma\omega},
\end{eqnarray*}
where
\begin{eqnarray}
&&iS_{U}[\hat{\Phi}^{\dagger}_{\sigma\omega},\hat{\Phi}_{\sigma\omega}] =
i S_{U}^{int}[\hat{\Phi}^{\dagger}_{\sigma\omega},\hat{\Phi}_{\sigma\omega}]\nonumber\\ &&+i\int_{-\infty}^{+\infty}\frac{d\omega}{2\pi}
\sum_{\sigma}\hat{\Phi}^{\dagger}_{\sigma\omega}(\omega + i(\Delta_{L} + \Delta_{R}))\hat{\sigma}_{3}\hat{\Phi}_{\sigma\omega}\label{eq_action}
\end{eqnarray}
is the effective action for a particle-hole symmetric ($E_{d} =0$) and interacting ($U \ne 0$) system
in the absence of an external magnetic field ($B=0$), and
\begin{eqnarray*}
i\Delta_{\lambda} = -\sum_{k,\sigma}\frac{|V_{k\lambda}|^{2}}{\omega - \epsilon_{k\lambda} + i\eta^{+}}\,\,\,\,\,\,\,\rm{for}\,\,\, \lambda = L, R
\end{eqnarray*}
is the coupling with the metallic leads, which in the limit
of a flat band ($\rho^{\lambda}(\omega) = \rho_{0}^{\lambda}$, $V_{k\lambda} = V_{\lambda}$)
of infinite bandwidth, tends to $i\Gamma_{\lambda} \rightarrow i\pi\rho_{0}^{\lambda}|V_{\lambda}|^{2}$.
In order to construct a perturbation theory
in the small parameters $E_{d}$, $B$, with respect to the reference system defined by the
action Eq.(\ref{eq_action}), let us introduce the dual fermion (Grassmann) fields $\hat{\phi}_{\sigma\omega}^{\dagger}=\left(f_{\sigma\omega}^{-},f_{\sigma\omega}^{+}\right)^{\dagger}$ where, as before, the index
$\mp$ refers to the time-ordered (anti-time-ordered) path along the
Keldysh contour. We insert the fermionic Hubbard-Stratonovich transformation,\cite{Rubtsov2008,Hafermann2009}
\begin{eqnarray*}
&&\int\mathcal{D}[\hat{\phi}_{\sigma\omega}^{\dagger},\hat{\phi}_{\sigma\omega}]\exp\left\{i\sum_{\sigma}\int_{-\infty}^{+\infty}
\frac{d\omega}{2\pi}\left[\hat{\phi}_{\sigma\omega}^{\dagger}\left(\mathbf{g}_{\sigma\omega}E_{d\sigma}
\hat{\sigma}_{3}\mathbf{g}_{\sigma\omega}\right)^{-1}\right.\right.\\
&&\left.\left.\times\hat{\phi}_{\sigma\omega} - \hat{\phi}_{\sigma\omega}^{\dagger}\mathbf{g}_{\sigma\omega}^{-1}\hat{\Phi}_{\sigma\omega} - \hat{\Phi}_{\sigma\omega}^{\dagger}\mathbf{g}_{\sigma\omega}^{-1}\hat{\phi}_{\sigma\omega}
\right]\right\}\nonumber\\
&&={\rm{Det}}\left[\left(\mathbf{g}_{\sigma\omega}E_{d\sigma}\hat{\sigma}_{3}\mathbf{g}_{\sigma\omega}\right)^{-1}\right]e^{-i\sum_{\sigma}\int_{-\infty}^{+\infty}\frac{d\omega}{2\pi}\hat{\Phi}_{\sigma\omega}^{\dagger}E_{d\sigma}\hat{\sigma}_{3}\hat{\Phi}_{\sigma,\omega}}
\end{eqnarray*}
into the partition function Eq.(\ref{eqB10}). Integrating out the local fermion field $\hat{\Phi}_{\sigma\omega}$,
one finds that the dual fermion bare Green's function is given by\cite{Munoz2013}
\begin{equation}
\mathbf{G}^{f(0)}_{\sigma\omega} = -\mathbf{g}_{\sigma\omega}\left(\mathbf{g}_{\sigma\omega} - E_{d\sigma}^{-1}\hat{\sigma}_{3} \right)^{-1}\mathbf{g}_{\sigma\omega}.
\label{eqB15}
\end{equation}
On the other hand, by functional differentiation of the partition function, an exact nonperturbative relation
between the dual fermion dressed Green's function $G^{f,ij}_{\sigma\omega}=-i\langle \hat{\phi}_{\sigma\omega}^{i\dagger}\hat{\phi}_{\sigma\omega}^{j} \rangle$ and
the local Green's function $G_{\sigma\omega}^{ij}=-i\langle \hat{\Phi}_{\sigma\omega}^{i\dagger} \hat{\Phi}_{\sigma\omega}^{j}\rangle$ is obtained\cite{Munoz2013}
\begin{equation}
\mathbf{G}_{\sigma,\omega} = -E_{d\sigma}^{-1}\hat{\sigma}_{3} +
\left(\mathbf{g}_{\sigma,\omega}E_{d\sigma}\hat{\sigma}_{3}\right)^{-1}
\mathbf{G}_{\sigma,\omega}^{f}\left(E_{d\sigma}\hat{\sigma}_{3}\mathbf{g}_{\sigma,\omega}\right)^{-1}
\label{eqB16}
\end{equation}
The dual fermion Green's function satisfies the matrix Dyson equation\cite{Munoz2013}
\begin{equation}
\mathbf{G}_{\sigma,\omega}^{f} = \mathbf{G}_{\sigma,\omega}^{f(0)} + \mathbf{G}_{\sigma,\omega}^{f(0)}\mathbf{\Sigma}^{f}_{\sigma\omega}\mathbf{G}_{\sigma,\omega}^{f}
\label{eqB17}
\end{equation}
Notice that the zeroth-order solution of Eqs.(\ref{eqB17},\ref{eqB16}) is
\begin{eqnarray*}
\mathbf{G}_{\sigma\omega}^{(0)} &=& -E_{d\sigma}\hat{\sigma}_{3} + \left(\mathbf{g}_{\sigma\omega}E_{d\sigma}\hat{\sigma}_{3} \right)^{-1}\mathbf{G}^{f(0)}_{\sigma\omega}
\left(E_{d\sigma}\hat{\sigma}_{3}\mathbf{g}_{\sigma\omega} \right)^{-1}\\
&=& \left(\mathbf{g}^{-1}_{\sigma\omega} - E_{d\sigma}\hat{\sigma}_{3} \right)^{-1}.
\end{eqnarray*}
The first-order solution for the Dyson equation (\ref{eqB17}) is
\begin{equation*}
\mathbf{G}_{\sigma,\omega}^{f(1)} = \mathbf{G}_{\sigma,\omega}^{f(0)}
+ \mathbf{G}_{\sigma,\omega}^{f(0)}\mathbf{\Sigma}_{\sigma,\omega}^{f}
\mathbf{G}_{\sigma,\omega}^{f(0)},
\end{equation*}
which upon substitution into Eq.(\ref{eqB16}) yields
\begin{eqnarray*}
\mathbf{G}_{\sigma,\omega}^{(1)}&=&\mathbf{G}_{\sigma,\omega}^{(0)} + \mathbf{G}_{\sigma,\omega}^{(0)}
\mathbf{\Sigma}^{f}_{\sigma,\omega}\mathbf{G}_{\sigma,\omega}^{(0)}\\
&=& [\mathbf{G}_{\sigma,\omega}^{(0)\,-1} -
\mathbf{\Sigma}^{f}_{\sigma,\omega}]^{-1} + O\left(\left[\mathbf{\Sigma}^{f}\right]^{2}\right)\\
&=& \left[\mathbf{g}_{\sigma,\omega}^{(0)\,-1}-\mathbf{\Sigma}_{\sigma,\omega}
-E_{d\sigma}\hat{\sigma}_{3} - \mathbf{\Sigma}^{f}_{\sigma,\omega} \right]^{-1}.
\end{eqnarray*}
It is clear then that, within this first-order solution of the Dyson equation, the perturbed
matrix self-energy is given by
\begin{equation}
\mathbf{\Sigma}_{\sigma,E_{d}}(\omega,B) = \left(E_{d}+\sigma b\right)\hat{\sigma}_{3}+\mathbf{\Sigma}_{\sigma,\omega} +
\mathbf{\Sigma}^{f}_{\sigma,\omega}.
\label{eqB21}
\end{equation}
\subsection{The dual Fermion self-energy}
In order to simplify the notation, let us use the multi-indexed labels $1\equiv (\omega_{1},\sigma_{1},i_{1})$ for frequency, spin and Keldysh contour $i_{1}=\mp$ indices. 
Let us define 
\begin{eqnarray*}
D_{12} &&\equiv (2\pi)\delta_{\omega_{1}-\omega_{2}}\delta_{\sigma_{1},\sigma_{2}}E_{d\sigma_{1}}[\hat{\sigma}_{3}]_{i_{1},i_{2}}\\
g_{12} &&\equiv (2\pi)\delta_{\omega_{1}-\omega_{2}}\delta_{\sigma_{1},\sigma_{2}}g_{\sigma_{1}\omega_{1}}^{i_{1},i_{2}}\\
\Gamma_{1234} &&\equiv (2\pi)\delta_{\omega_{1}+\omega_{3}-\omega_{2}-\omega_{4}}\left[\Gamma_{\sigma_{1}\sigma_{2};\sigma_{3}\sigma_{4}}(\omega_{1},\omega_{2};\omega_{3},\omega_{4}) \right]^{i_{1},i_{2};i_{3},i_{4}}.
\end{eqnarray*}
Here, $\Gamma_{\sigma_{1}\sigma_{2};\sigma_{3}\sigma_{4}}(\omega_{1},\omega_{2};\omega_{3},\omega_{4})$
is the four-point vertex of the reference system.
The dual fermion self-energy is given by the expression
\begin{eqnarray*}
\Sigma^{f}_{12} &\equiv & (2\pi)\delta_{\omega_{1}-\omega_{2}}\delta_{\sigma_{1},\sigma_{2}}\left[\Sigma^{f}_{\sigma_{1},\omega_{1}} \right]_{i_{1},i_{2}}\\
&=& i \Gamma_{1234}g_{44'}
\left[g - D^{-1} \right]^{-1}_{4'3'}g_{3'3},
\end{eqnarray*}
where in this context repeated indices stand for a generalized convolution in frequency, spin and
Keldysh-contour indices. A  series
expansion of the dual fermion self-energy matrix follows from
$\left[g - D^{-1} \right]^{-1}=-D\left[I - gD \right]^{-1} = -D - DgD -DgDgD + \ldots$:
\begin{eqnarray}
\Sigma^{f}_{12} = -i\Gamma_{1234}\left[gDg \right]_{43} 
- i\Gamma_{1234}\left[gDgDg \right]_{43} + O(D^{3})\nonumber\\
\label{eqB24}
\end{eqnarray}
The four-point vertex is given by
\begin{eqnarray*}
\Gamma_{\sigma\sigma';\sigma'\sigma}^{(0)----} &=& \Gamma_{0}(1 - \delta_{\sigma,\sigma'})\\
\Gamma_{\sigma\sigma';\sigma'\sigma}^{(0)++++} &=& -\Gamma_{0}(1 - \delta_{\sigma,\sigma'}),
\end{eqnarray*}
where 
\begin{equation}
\Gamma_{0}(U) = U + \pi\Delta\left(15 - 3\pi^{2}/2 \right)(U/\pi\Delta)^{3} + O(U^{5}).
\label{eqvertex}
\end{equation}
The first term on the right-hand side of Eq.(\ref{eqB24}) possesses only two non-vanishing diagonal matrix
elements,
\begin{eqnarray*}
&&-i\sum_{\sigma',j=\pm}\int\frac{d\omega'}{2\pi}\Gamma_{\sigma\sigma';\sigma'\sigma}^{(0)----}g_{\sigma'\omega'}^{-j}E_{d\sigma'}\left[\hat{\sigma}_{3} \right]_{jj}g_{\sigma'\omega'}^{j-}\\
&&=-i\Gamma_{0}E_{d,-\sigma}Z_{-\sigma}^{--},
\end{eqnarray*}
and $+i\Gamma_{0}E_{d,-\sigma}Z_{-\sigma}^{++}$, where we have defined
\begin{eqnarray*}
Z_{\sigma}^{--} = \int_{-\infty}^{+\infty}\frac{d\omega'}{2\pi}\left(\left[g_{-\sigma\omega'}^{--} \right]^2
-g_{-\sigma\omega'}^{-+}g_{-\sigma\omega'}^{+-} \right)
\end{eqnarray*}
and $Z_{\sigma}^{++} = \left(Z_{\sigma}^{--} \right)^{*}$.
Direct calculation of the integral, and consistently with the approximation
for the reference system keeping terms up to $O(\Gamma_{0}^{2})$ only, we obtain
\begin{eqnarray}
\Sigma^{f,--}_{\sigma}(\omega,B)&=&-i\Gamma_{0}E_{d,-\sigma}Z_{-\sigma}^{--}\nonumber\\
&=& -\left(E_{d} + \sigma b \right)\tilde{u}\left\{1-\frac{1}{3}\left[\left(\frac{\pi T}{\tilde{\Delta}} \right)^{2} + \left(\frac{eV}{\tilde{\Delta}} \right)^{2} \right]\right.\nonumber\\
&&\left.+\frac{7}{9}\zeta\left(\frac{\pi T e V}{\tilde{\Delta}^{2}} \right)^{2} \right\},
\label{eqB28}
\end{eqnarray}
with the other components given by $\Sigma^{f,++}_{\sigma}(\omega,B) = -\Sigma^{f,--}_{\sigma}(\omega,B)$, $\Sigma^{f,+-}_{\sigma}(\omega,B) = \Sigma^{f,-+}_{\sigma}(\omega,B)=0$.

\subsection{The retarded self-energy}
\label{subsec:retarded self-energy}
At the order of approximation of Eq.(\ref{eqB21}) and Eq.(\ref{eqB28}) \cite{Munoz2013}, the self-energy components
at the local site are
\begin{eqnarray}
\Sigma_{\sigma,E_{d}}^{++}(\omega,B) &=& \Sigma^{++}_{\sigma\omega} - E_{d\sigma} + \Sigma^{f,++}_{\sigma}(\omega,B)\nonumber\\
\Sigma_{\sigma,E_{d}}^{--}(\omega,B) &=& \Sigma^{--}_{\sigma}(\omega,B) + E_{d\sigma} + \Sigma^{f,--}_{\sigma}(\omega,B)\nonumber\\
\Sigma_{\sigma,E_{d}}^{+-}(\omega,B) &=& \Sigma^{+-}_{\sigma}(\omega,B)\label{eqB29}\\
\Sigma_{\sigma,E_{d}}^{-+}(\omega,B) &=& \Sigma^{-+}_{\sigma}(\omega,B).\nonumber
\end{eqnarray}
We thus obtain the retarded self-energy from the relation $\Sigma_{\sigma,E_{d}}^{r}(\omega,B) = \Sigma_{\sigma,E_{d}}^{+-}(\omega,B) - \Sigma_{\sigma,E_{d}}^{++}(\omega,B)$, as follows
\begin{eqnarray}
&&\Sigma_{\sigma,E_{d}}^{r}(\omega,B) = (1-\tilde{\chi}_{++})\omega + E_{d}-\sigma b - (E_{d}+\sigma h)\tilde{u}\nonumber\\
&&\times\left\{1
-\frac{1}{3}
\left[\left(\frac{\pi T}{\tilde{\Delta}} \right)^{2} + \left(\frac{eV}{\tilde{\Delta}} \right)^{2} \right]+\frac{7}{9}\zeta\left(\frac{\pi T eV}{\tilde{\Delta}^{2}} \right)^{2}\right\}\nonumber\\
&&+ i\Delta\tilde{u}^{2}\left[\left(\frac{\omega}{\tilde{\Delta}} \right)^{2} + \left(\frac{\pi T}{\tilde{\Delta}} \right)^{2} + \zeta\left(\frac{eV}{\tilde{\Delta}} \right)^{2}-\frac{\zeta}{3}\left(\frac{\pi T eV}{\tilde{\Delta}^{2}} \right)^{2} \right]\nonumber\\
\label{eqB30}
\end{eqnarray}
Here, $\tilde{u} = z\Gamma_{0}/(\pi\Delta)$ is the renormalized interaction, for $z=\tilde{\chi}_{++}^{-1}$
the wave function renormalization factor for the particle-hole symmetric reference system at zero magnetic field.
The renormalized quasiparticle spectral broadening is $\tilde{\Delta}=z\Delta$.
The retarded Green's function corresponding to this self-energy is 
\begin{eqnarray*}
G_{\sigma,E_{d}}^{r}(\omega,B) &=& \left(\omega + i\Delta - \Sigma_{\sigma,E_{d}}^{r}(\omega,B) \right)^{-1}\\
&=& \tilde{\chi}_{++}^{-1}\left(\omega - \tilde{E}_{d} + \sigma \tilde{b} + i\tilde{\Delta} - \tilde{\Sigma}^{r}_{\sigma\omega}(B) \right)^{-1}
\end{eqnarray*}
Here, we have defined the renormalized self-energy
\begin{eqnarray*}
&&\tilde{\Sigma}^{r}_{\sigma\omega}(B) = - (\tilde{E}_{d}+\sigma \tilde{b})\tilde{u}
\left\{1-\frac{1}{3}
\left[\left(\frac{\pi T}{\tilde{\Delta}} \right)^{2}\right.\right.\\
&&\left.\left. + \left(\frac{eV}{\tilde{\Delta}} \right)^{2} \right]+\frac{7}{9}\zeta\left(\frac{\pi T eV}{\tilde{\Delta}^{2}} \right)^{2}\right\}\\
&&+ i\tilde{\Delta}\tilde{u}^{2}\left[\left(\frac{\omega}{\tilde{\Delta}} \right)^{2} + \left(\frac{\pi T}{\tilde{\Delta}} \right)^{2} + \zeta\left(\frac{eV}{\tilde{\Delta}} \right)^{2}-\frac{\zeta}{3}\left(\frac{\pi T eV}{\tilde{\Delta}^{2}} \right)^{2} \right]
\end{eqnarray*}

In the above, we have systematically included all contributions
up to second order in the renormalized Coulomb interaction $\tilde{u}$
and particle-hole asymmetry $\tilde{\varepsilon}_{d}=\tilde{E_{d}}/\tilde{\Delta}=E_{d}/\Delta$. 
This corresponds to approximating the dual fermion Green's function by
\begin{eqnarray}
\mathbf{G}^{f} = \mathbf{G}_{0}^{f} + \mathbf{G}_{0}^{f}\mathbf{\Sigma}^{f}\mathbf{G}_{0}^{f}
\end{eqnarray}
instead of a selfconsistent solution of the Dyson equation $\mathbf{G}^f=\mathbf{G}_0^f+\mathbf{G}_0^f\mathbf{\Sigma}^f\mathbf{G}^f$.
The self-energy in this equation involves a single
renormalized four-point vertex, as stated by Eq.~(\ref{eqB24}).
Additional contributions to the self-energy are generated by including reducible contributions to the four-point vertex.
This involves the  entire family of "parquet" diagrams. Explicit calculations of these higher-order contributions are currently
under development, but go beyond the scope of this paper.

\subsection{Differential conductance $G(T,B)$}
\label{subsec:differential conductance}
The differential conductance $G(T,B) \equiv dI/dV$ is expressed by the formula
\begin{eqnarray*}
G(T,B) = \frac{e^2}{h}\int_{-\infty}^{+\infty}d\omega
\left(-\frac{\partial f}{\partial\omega} \right)\sum_{\sigma}\mathcal{T}_{\sigma}(\omega,T,B)
\end{eqnarray*}
Here, the transmission is
\begin{eqnarray*}
\mathcal{T}_{\sigma}(\omega,T,B) = 4\pi\frac{\Delta_{L}\Delta_{R}}{\Delta_{L}+\Delta_{R}}A_{\sigma}(\omega,T,B),
\end{eqnarray*}
where the spectral function is defined as
\begin{eqnarray*}
A_{\sigma}(\omega,T,B) = -\frac{1}{\pi}{\rm{Im}}G_{\sigma,E_{d}}^{r}(\omega,B).
\end{eqnarray*}
The differential conductance at zero bias, and up to second order in temperature and magnetic field can be cast into the form
\begin{equation}
\frac{G(T,B)}{G_{0}} = 1 - c'_{T}\left(\frac{T}{T_{0}^{(s)}} \right)^{2} - c'_{B}\left(\frac{g\mu_{B}B/2}{k_{B}T_{0}^{(s)}}\right)^{2}
\label{eqBconductance}
\end{equation} 
Here, the Kondo scale is based on the spin susceptibility of the particle-hole symmetric system 
$\chi^{s}(0)=(g\mu_{B})^2/4T_{0}^{(s)}$, with
\begin{eqnarray*}
\chi^{s}(0) &=& \frac{(g\mu_{B})^{2}}{2}\tilde{A}_{d}(0)\left(1 + \tilde{U}\tilde{A}_{d}(0) \right)\nonumber\\
&=&\frac{(g\mu_{B})^{2}}{2}\frac{1}{\pi\tilde{\Delta}}\left(1 + \frac{\tilde{U}}{\pi\tilde{\Delta}}\right)\nonumber\\
&=& \frac{(g\mu_{B})^{2}}{2}\frac{1}{\pi\tilde{\Delta}}(1 + \tilde{u}),\label{eqBchi}
\end{eqnarray*}
and we have used $\tilde{A}_{d}(0) = 1/(\pi\tilde{\Delta})$. Thus,
\begin{equation}
T_{0}^{(s)} =\frac{\pi\tilde{\Delta}}{2(1+\tilde{u})}.\label{eq:spt-t0}
\end{equation}
The coefficients in Eq.(\ref{eqBconductance})are given by
\begin{eqnarray*}
c'_{T} &=& \frac{\pi^{4}}{12}\frac{1+2\tilde{u}^2 +(1-\tilde{u})(5\tilde{u}-3)\tilde{\varepsilon}_{d}^{2}}{(1+\tilde{u})^{2}(1+(1-\tilde{u})^{2}\tilde{\varepsilon}_{d}^{2})^2}\\
c'_{B} &=& \frac{\pi^{2}}{16}\frac{1-3(1-\tilde{u})^{2}\tilde{\varepsilon}_{d}^{2}}{(1+(1-\tilde{u})^2\tilde{\varepsilon}_{d}^{2})^2},
\end{eqnarray*}
where we have set ${\tilde{\varepsilon}}_{d} = \tilde{E}_{d}/\tilde{\Delta} = (\varepsilon_{d}+U/2)/\Delta$.

\subsection{Derivation of the relation $\tilde{u}=R-1$}
From Fermi liquid theory, we have the general result
\begin{equation}
\tilde{A}_{d,\sigma}(0) = z^{-1}A_{d,\sigma}(0)=\frac{\sin^{2}(\pi n_{d\sigma})}{\pi\tilde{\Delta}}
\label{eqRR1}
\end{equation}
where $z$ is the wave function renormalization factor, and $n_{d\sigma}$ is the local level occupancy for spin $\sigma$. 
Along with this, we have the following Fermi
liquid relations for the specific heat, spin and charge susceptibilities \cite{Hewson1993}
\begin{eqnarray}
\gamma_{d} &=& \frac{2\pi^{2}k_{B}^{2}}{3}\tilde{A}_{d,\sigma}(0)\label{eqRR2}\\
\chi_{d} &=& \frac{(g\mu_{B})^{2}}{2}\tilde{A}_{d,\sigma}(0)\left(1 + \tilde{U}\tilde{A}_{d,\sigma}(0) \right)\label{eqRR2b}\\
\chi_{c,d} &=& 2\tilde{A}_{d,\sigma}(0)\left(1 - \tilde{U}\tilde{A}_{d,\sigma}(0) \right)\label{eqRR2c}
\end{eqnarray}
where $\tilde{U} = z^{2}\Gamma_{0}$, $\tilde{\Delta} = z\Delta$.
From Eqs.~(\ref{eqRR2})--(\ref{eqRR2c}), we obtain
\begin{eqnarray}
\frac{4}{(g\mu_{B})^{2}}\chi_{d} + \chi_{c,d} = \frac{6}{\pi^{2}k_{B}^{2}}\gamma_{d},
\label{eqRR3}
\end{eqnarray}
and together with the definition of the Wilson ratio, combined with Eqs.~(\ref{eqRR2})--(\ref{eqRR3}), we obtain
\begin{equation}
R \equiv \frac{4\pi^{2}k_{B}^{2}}{3(g\mu_{B})^{2}}\frac{\chi_{d}}{\gamma_{d}} = 1 + \tilde{U}\tilde{A}_{d,\sigma}(0)
\label{eqRR4}
\end{equation}
Substituting Eq.(\ref{eqRR1}) into Eq.(\ref{eqRR4}), we obtain
\begin{equation*}
R = 1 + \frac{\tilde{U}}{\pi\tilde{\Delta}}\sin^{2}(\pi n_{d\sigma}).
\end{equation*}
Let us define $\tilde{u} \equiv \tilde{U}/(\pi\tilde{\Delta}) = z \Gamma_{0}/(\pi\Delta)$, with
$\Gamma_{0}$ defined by Eq.(\ref{eqvertex}) as the four-point vertex. Then, we
have
\begin{equation}
\tilde{u} = \frac{R-1}{\sin^{2}(\pi n_{d\sigma} )}.\label{eqRR5}
\end{equation}
Finally, notice that for a particle-hole symmetric system  $n_{d\sigma} = 1/2$, and hence $\tilde{u} = R - 1$.
\bibliography{scaling}
\end{document}